\UseRawInputEncoding
\documentclass[pra,onecolumn]{revtex4}
\usepackage{amsthm}
\usepackage{amsmath}
\usepackage{latexsym}
\usepackage{amsfonts}
\usepackage{amssymb}
\usepackage{color}
\usepackage{bbm,dsfont}
\usepackage{graphicx}
\usepackage{hyperref}
\usepackage{subfigure}
\usepackage{caption}
\usepackage{xr}
\usepackage{diagbox}
\usepackage{booktabs}
\newtheorem{proposition?}{Proposition?}

\theoremstyle{definition}

\begin{document}
\title[]{Detecting Tripartite Steering via Quantum Entanglement}
\author{Zhihua Chen}
\affiliation{School of Science, Jimei University, Xiamen 361021,China}

%\author{Zhihua Chen}
%\affiliation{School of Science, Jimei University, Xiamen 361021, China}
\author{Shao-Ming Fei}
\thanks{Correspondence:  feishm@cnu.edu.cn}
\affiliation{School of Mathematical Sciences, Capital Normal University, Beijing 100048, China,\\
Max Planck Institute for Mathematics in the Sciences, 04103 Leipzig, Germany}

% The commands \thirdnote{} till \eighthnote{} are available for further notes

%\simplesumm{} % Simple summary

%\conference{} % An extended version of a conference paper

% Abstract (Do not insert blank lines, i.e. \\)

\begin{abstract}Einstein-Podolsky-Rosen steering is a kind of powerful nonlocal quantum resource in quantum information processing such as quantum cryptography and quantum communication. Many criteria have been proposed in the past few years to detect steerability, both analytically and numerically, for bipartite quantum systems. We propose effective criteria for tripartite steerability and genuine tripartite steerability of three-qubit quantum states by establishing connections between the tripartite steerability (resp. genuine tripartite steerability) and the tripartite entanglement (resp. genuine tripartite entanglement) of certain corresponding quantum states.
From these connections, tripartite steerability and genuine tripartite steerability can be detected without using any steering inequalities. The ``complex cost'' of determining tripartite steering and genuine tripartite steering can be reduced by detecting the entanglement of the newly constructed states in the experiment. Detailed examples are given to illustrate the power of our criteria in detecting the (genuine) tripartite steerability of tripartite states.

\
\

\end{abstract}

% Keywords

\maketitle

\section{Introduction}

Originally introduced by Schr$\ddot{o}$dinger \cite{Schr} the Einstein-Podolsky-Rosen (EPR) steering for bipartite systems was considered as a 'spooky action at distance' \cite{Eins} in the sense that one party can steer another distant party's state instantly. The concept of EPR steering was proposed by Wiseman, Jones, and Doherty in 2007 \cite{Wiseman}. Since then the EPR steering has been systematically studied. Many different methods were proposed to detect and quantify the steerability of bipartite quantum states \cite{ineq1,ineq2,ineq3,ineq4,ineq5,ineq6,unc1,unc2,unc3,mom,A-V-N,SDP}, together with many applications in quantum information processing tasks including one-sided device-independent quantum key distribution, random generation and one-sided device-independent quantum self-testing of pure quantum states, subchannel discrimination, quantum communication \mbox{et al. \cite{Branciard, Passaro, Coyle,Piani,Sun,Xiang,He1,Ku,Chiu,Chen1}.}

The EPR steering lies between quantum nonlocality and quantum entanglement.
A bipartite state is quantum nonlocal if it does not admit a local hidden variable model \cite{Bell}, while it is EPR steerable if it does not admit a hidden state model \cite{Wiseman}.

Bipartite steering is defined as follows. Alice and Bob share a quantum state $\rho_{\mathcal{AB}}$. Alice performs black-box measurements $A$ with outcomes $a$, denoted by $M_A^a$ ($M_A^a\geq 0$ $\forall A, a$ and $\sum\limits_a M_A^a={\rm{I}}$ $\forall A$, with $\rm{I}$ denoting the identity operator). The set of unnormalized conditional states $\{\delta_{A}^a\}$ on Bob's side is called an assemblage. Each element in this assemblage is given~by
\begin{equation}
\begin{aligned}
\delta_{A}^a={\rm{Tr}}[(M_A^a\otimes{\rm{I}}).\rho_{\mathcal{AB}}].
\end{aligned}
\end{equation}

Alice can not steer Bob if $\delta_{A}^a$ admits a local hidden state model (LHS), i.e., $\delta_{A}^a$ admits the decomposition
\begin{equation}
\begin{aligned}
\delta_{A}^a=\sum\limits_{\lambda}p(\lambda)p(a|A,\lambda)\rho_{\lambda}^{\beta},
\end{aligned}
\end{equation}
where $\lambda$ denotes classical random variable which occurs with
probability $p(\lambda)$ satisfying $\sum\limits_{\lambda}p(\lambda)=1,$ $p(a|A,\lambda)$ is the probability given by the black-box measurement on Alice's side, $\rho_{\lambda}^{\beta}$ are some local hidden states. Bob performs measurement $B$ with outcomes $b$, denoted by $M_B^b,$ on the assemblage. The joint probability is $p(a,b|A,B)={\rm{Tr}}[M_B^b\delta_A^a]$.
$\rho_{\mathcal{AB}}$ is said to be a steerable state from Alice to Bob if $p(a,b|A,B)$ does not admit a local hidden variable-local hidden state (LHV-LHS) model of the form,
\begin{equation}
\begin{aligned}
p(a,b|A,B)=\sum\limits_{\lambda}p(\lambda)p(a|A,\lambda)p_Q(b|B,\rho_{\lambda}^{\beta}).
\end{aligned}
\end{equation}

Different from quantum entanglement and quantum nonlocality, EPR steering is asymmetric in general, which means that Alice can steer Bob but not vice versa for some bipartite quantum states $\rho_{\mathcal{AB}}$ \cite{Bowles}. The bipartite quantum nonlocality and EPR steering can be detected by detecting the EPR steering and quantum entanglement of some newly constructed quantum states, respectively \cite{Chen, Das, Lai}.

The multipartite steering is an important resource in quantum communication networks \cite{He,Armstrong,Li} and in one-sided or two-sided device-independent entanglement
detections~\cite{Cavalcantin,Jebaratnam}.
Some ambiguities exist in the definition of multipartite
steering. With respect to the typical spooky action at a distance \cite{He,Armstrong,Li}, and the semi-device independent entanglement verification scheme \cite{Cavalcantin,Jebaratnam}, two different approaches have been introduced to define the multipartite steering
\cite{He, Cavalcantin,Jebaratnam}.
One approach is to define genuine multipartite steering in terms of the steering under bi-partitions. A tripartite state $\rho_{\mathcal{ABC}}$ is defined to be genuine tripartite steerable if the state does not admit the mixtures of bi-partitions where in each partition (e.g., $\mathcal{A}|\mathcal{BC}$) the two-party state (e.g., $\mathcal{BC}$) is allowed to be steerable. Linear inequalities have been derived to detect this kind of genuine multipartite steering \cite{He} and used in experimental demonstrations \cite{Armstrong, Li}.

Another approach to defining tripartite steering and genuine tripartite steering is given as follows \cite{Cavalcantin,Jebaratnam}. Let $p(a,b,c|A,B,C)$ be the joint probability that Alice, Bob and Charlie perform measurements $A$, $B$ and $C$ with outcomes $a,$ $b$ and $c$, given by measurements operators $M_A^a$, $M_B^b$ and $M_C^c$, respectively.
A quantum state $\rho_{\mathcal{ABC}}$ is said to be tripartite steerable from Alice (untrusted party) to Bob and Charlie (trusted parties) if $p(a,b,c|A,B,C)$ does not admit a fully LHV-LHS model such that
\begin{equation}\label{MS-A-BC}
\begin{aligned}
p(a,b,c|A,B,C)
=\sum\limits_{\lambda} p(\lambda) p(a|A,\lambda)p_Q(b|B,\tau^{\beta}_{\lambda})p_Q(c|C,\tau^{\gamma}_{\lambda}),
\end{aligned}
\end{equation}
where $p_Q(b|B,\tau^{\beta}_{\lambda})={\rm{Tr}}[M_B^b\tau^{\beta}_{\lambda}]$ and $p_Q(c|C,\tau^{\gamma}_{\lambda})={\rm{Tr}}[M_C^c\tau^{\gamma}_{\lambda}]$ are the distributions from the local
hidden states $\tau_{\lambda}^{\beta}$ and $\tau_{\lambda}^{\gamma}$, see Equation~(13) in \cite{Jebaratnam} and Equation~(2) in \cite{Riccardi}.

The genuine tripartite steering has been defined in \cite{Cavalcantin,Jebaratnam,Riccardi}. Alice measures her system so as to nonlocally influence the state of the other two parties. The ensemble of the unnormalized states is given by
\begin{equation}\label{ensa-bc}
\begin{aligned}
\{\delta_{M_A^a}^{BC}={\rm{Tr}}[(M_A^a\otimes {\rm{I}}\otimes{\rm{I}}).\rho_{\mathcal{ABC}}]\}.
\end{aligned}
\end{equation}

{If the ensemble prepared on Bob's and Charlie's sides cannot be reproduced by a
biseparable state as Equation~(\ref{bi}),}
{\begin{equation}\label{bi}
\begin{aligned}
\rho_{\mathcal{ABC}}=&\sum\limits_{\lambda}p_1(\lambda)\rho_{\lambda}^{\alpha}\otimes
\rho_{\lambda}^{\beta\gamma}+\sum\limits_{\lambda}p_2(\lambda)\rho_{\lambda}^{\alpha\beta}
\otimes\rho_{\lambda}^{\gamma}+\sum\limits_{\lambda}p_3(\lambda)\rho_{\lambda}^{\beta}\otimes \rho_{\lambda}^{\alpha\gamma},
\end{aligned}
\end{equation}
with $\sum\limits_{\lambda}p_1(\lambda)+p_2(\lambda)+p_3(\lambda)=1,$
then $\rho_{\mathcal{ABC}}$ is not genuine tripartite steerable
from Alice to Bob and Charlie. Therefore, if $\rho_{\mathcal{ABC}}$ is genuine tripartite steerable from Alice to Bob and Charlie, then}
each member of the ensemble (\ref{ensa-bc}) can not be expressed as \cite{Cavalcantin,Riccardi},
\begin{equation}\label{8}
\begin{aligned}
\delta_{M_A^a}^{BC}=&\sum\limits_{\lambda}p_1(\lambda)p_1(a|A,\lambda)
\rho_{\lambda}^{\beta\gamma}\\
&+\sum\limits_{\lambda}p_2(\lambda)\delta_{M_A^a,\lambda}^{\beta}
\otimes\rho_{\lambda}^{\gamma}\\
&+\sum\limits_{\lambda}p_3(\lambda)\rho_{\lambda}^{\beta}\otimes \delta_{M_A^a,\lambda}^{\gamma}
\end{aligned}
\end{equation}
with $\delta_{M_A^a,\lambda}^{\beta}={\rm{Tr}}_{\mathcal{A}}[(M_A^a\otimes{\rm{I}})
\rho_{\lambda}^{\alpha\beta}]$
and  $\delta_{M_A^a,\lambda}^{\gamma}={\rm{Tr}_{\mathcal{A}}}[(M_A^a\otimes{\rm{I}})
\rho_{\lambda}^{\alpha\gamma}].$
The first term on the right-hand side of (\ref{8}) stands for that Alice cannot steer Bob and Charlie. Bob and Charlie share entanglement and a local hidden entangled
state $\rho_{\lambda}^{\beta\gamma}$. The other two terms imply that there is no entanglement between Bob and Charlie, and Alice can steer one of the two systems but not both: the second (third) term stands for that
Alice can steer Bob (Charlie) but not Charlie (Bob).

A state is genuine tripartite steerable from Alice to Bob and Charlie
if the joint probability $p(a,b,c|A,B,C)={\rm{Tr}}[(M_B^b\otimes M_C^c)\delta_{M_A^a}^{BC}]$ does not admit a hybrid LHV-LHS model~\cite{Jebaratnam,Riccardi},
\begin{equation}\label{1pp}
\begin{aligned}
&p(a,b,c|A,B,C)
=\sum\limits_{\lambda} p_1(\lambda) p(a|A,\lambda)p_Q(b,c|B,C,\rho_{\lambda}^{\beta\gamma})\\
&\hspace{2.5cm}+\sum\limits_{\lambda} p_2(\lambda) p_Q(a,b|A,B)p_Q(c|C,\rho_{\lambda}^{\gamma})\\
&\hspace{2.5cm}+\sum\limits_{\lambda} p_3(\lambda) p_Q(a,c|A,C)p_Q(b|B,\rho_{\lambda}^{\beta}),
\end{aligned}
\end{equation}
where $p(a|A,\lambda)$ is the distribution on Alice's side from black-box measurements performed on a  quantum state, $p_Q(c|C,\rho_{\lambda}^{\gamma})$ and $p_Q(b|B,\rho_{\lambda}^{\beta})$ are the distributions from measurements on quantum states $\rho_{\lambda}^{\gamma}$ and $\rho_{\lambda}^{\beta}.$ $p_Q(b,c|B,C,\rho_{\lambda}^{\beta\gamma})$ can be reproduced by  quantum state $\rho_{\lambda}^{\beta\gamma}$ shared by Bob and Charlie. $p_Q(a,b|A,B)={\rm{Tr}}[(M_A^a\otimes M_B^b)\rho_{\lambda}^{\alpha\beta}]$ and $p_Q(a,c|A,C)={\rm{Tr}}[(M_A^a\otimes M_C^c)\rho_{\lambda}^{\alpha\gamma}]$ are distributions from a quantum state with untrusted $\mathcal{A}$ and trusted $\mathcal{B}$ and $\mathcal{C}$.
When $M_A^a=\rm{I},$ $p_Q(b|B)$ and $p_Q(c|C)$ are probabilities from the local hidden states ${\rm{Tr}}_{\mathcal{A}}[\rho_{\lambda}^{\alpha\beta}]$ and ${\rm{Tr}}_{\mathcal{A}}[\rho_{\lambda}^{\alpha\gamma}]$, respectively, since $\mathcal{B}$ and $\mathcal{C}$ are the trusted parties. We always use $p_Q(x,y|X,Y)$ $(x,y =a,b$ or $c$ and $X, Y=A, B$ or $C)$ to represent the distribution from measurements on two parties with one party trusted and the other two untrusted in this~paper.

A quantum state $\rho_{\mathcal{ABC}}$ is said to be tripartite steerable from (untrusted) Alice and Bob to (trusted) Charlie if the joint probability $p(a,b,c|A,B,C)$ does not admit a fully LHV-LHS model such that
\begin{equation}\label{MS-AB-C}
\begin{aligned}
p(a,b,c|A,B,C)
=\sum\limits_{\lambda} p(\lambda) p(a|A,\lambda)p(b|B,\lambda)p_Q(c|C,\tau_{\lambda}^{\gamma}),
\end{aligned}
\end{equation}
where $p(a|A,\lambda)$ and $p(b|B,\lambda)$ are the probabilities from the black-box measurements, $p_Q(c|C,\tau_{\lambda}^{\gamma})$ is the distribution from local hidden state $\tau_{\lambda}^{\gamma}$, see also the definition given in~\cite{Jebaratnam,Riccardi}.

The genuine tripartite steering from Alice and Bob to Charlie has also been defined in \cite{Jebaratnam,Cavalcantin,Riccardi}. Alice and Bob measure their systems so as to nonlocally influence the state of Charlie's. The ensemble prepared on  Charlie's side cannot be reproduced by a biseparable state as Equation~(\ref{bi}). Each member in the ensemble of unnormalized states can not be given by
\begin{equation}\label{ensab-c}
\begin{aligned}
\delta_{M_A^a,M_B^b}^{C}=&{\rm{Tr}}[(M_A^a\otimes M_B^b\otimes{\rm{I}}).\rho_{\mathcal{ABC}}]\\
=&\sum\limits_{\lambda}p_1(\lambda)p_1(a,b|A,B,\lambda)\rho_{\lambda}^{\gamma}\\
&+\sum\limits_{\lambda}p_2(\lambda)p(a|A,\lambda)\delta_{M_B^b,\lambda}^{\gamma}\\
&+\sum\limits_{\lambda}p_3(\lambda)p(b|B,\lambda)\delta_{M_A^a,\lambda}^{\gamma},
\end{aligned}
\end{equation}
with $\delta_{M_B^b,\lambda}^{\gamma}={\rm{Tr}_{\mathcal{B}}}[(M_B^b\otimes{\rm{I}})
\rho_{\lambda}^{\beta\gamma}]$ and  $\delta_{M_A^b,\lambda}^{\gamma}={\rm{Tr}_{\mathcal{A}}}[(M_A^a\otimes{\rm{I}})
\rho_{\lambda}^{\alpha\gamma}]$. The first term on the right-hand side of (\ref{ensab-c}) stands for that  Alice and Bob cannot jointly steer Charlie, and the second (third) term stands for that
only Bob (Alice) can steer the state of Charlie.
A state is genuine tripartite steerable from Alice and Bob to Charlie
if the joint probability $p(a,b,c|A,B,C)$ does not admit a hybrid LHV-LHS model such that
\begin{equation}\label{GMS-AB-Cp}
\begin{aligned}
&p(a,b,c|A,B,C)=\sum\limits_{\lambda} p_1(\lambda) p(a,b|A,B,\lambda)p_Q(c|C,\rho_{\lambda}^{\gamma})\\
&\hspace{2.5cm}+\sum\limits_{\lambda} p_2(\lambda) p(a|A,\lambda)p_Q(b,c|B,C)\\
&\hspace{2.5cm}+\sum\limits_{\lambda} p_3(\lambda) p(b|B,\lambda)p_Q(a,c|A,C),
\end{aligned}
\end{equation}
where $\sum\limits_{\lambda}p_1(\lambda)+\sum\limits_{\lambda}p_2(\lambda)
+\sum\limits_{\lambda}p_3(\lambda)=1.$  $p(a|A,\lambda)$ and $p(b|B,\lambda)$ are the distributions on Alice's and Bob's sides, respectively, arising from black-box measurements performed on a  quantum state.
$p(a,b|A,B,\lambda)$ is the distribution produced from black-box measurements performed on a  quantum state. $p_Q(c|C,\rho_{\lambda}^{\gamma})$ is the distribution from the state $\rho_{\lambda}^{\gamma}$. $p_Q(b,c|B,C)={\rm{Tr}[(M_B^b\otimes M_C^c)}\rho_{\lambda}^{\beta\gamma}]$ and $p_Q(a,c|A,C)={\rm{Tr}[(M_A^a\otimes M_C^c)}\rho_{\lambda}^{\alpha\gamma}]$ are probabilities from a $2$-qubit quantum state with untrusted $\mathcal{A}$ and $\mathcal{B}$ and trusted $\mathcal{C}$. When $M_A^a=M_B^b=\rm{I},$ $p_Q(c|C)$ are probabilities from the local hidden states ${\rm{Tr}}_{\mathcal{B}}[\rho_{\lambda}^{\beta\gamma}]$ and ${\rm{Tr}}_{\mathcal{A}}[\rho_{\lambda}^{\alpha\gamma}]$, respectively, since $\mathcal{C}$ is the trusted party.

Entropic steering inequalities and semi-definite-program have been adopted to investigate the detection of multipartite steering \cite{Cavalcantin,Riccardi,Costa}. In the following, we construct new quantum states with respect to given three-qubit states and detect the tripartite steering and genuine tripartite steering analytically in terms of the tripartite entanglement and the genuine tripartite entanglement of the newly constructed quantum states.
The entanglement of the newly constructed states can be detected by using the entanglement witness without full tomography of the states. By detecting the entanglement of the newly constructed states, the tripartite steering and genuine tripartite steering can be detected without using any steering inequalities. Since the ``complexity cost'' (the number of possible patterns of joint detection outcomes that can occur, see \cite{Saunders}) for the least complex demonstration of entanglement is less than the ``complexity cost'' for the least complex demonstration of EPR steering \cite{Das,Saunders}, our scheme reduces the ``complex cost'' in experimental steering demonstration.

\section{Main Results}
A quantum state is fully separable if the joint probability $p(a,b,c|A,B,C)$ satisfies the~condition,
\begin{equation}\label{fulls}
\begin{aligned}
p(a,b,c|A,B,C)
=\sum\limits_{\lambda}p_{\lambda}p_Q(a|A,\tau_{\lambda}^{\alpha})p_Q(b|B,\tau_{\lambda}^{\beta})
p_{Q}(c|C,\tau_{\lambda}^{\gamma}).
\end{aligned}
\end{equation}

Fully separable states are neither tripartite steerable states from Alice to Bob and Charlie nor from Alice and Bob to Charlie.
From (\ref{MS-A-BC}) and (\ref{MS-AB-C}) a state which is not tripartitely steerable from Alice to Bob and Charlie is not tripartitely steerable from Alice and Bob to Charlie,
i.e., tripartite steering from Alice and Bob to Charlie
is stronger than that from Alice to Bob and Charlie.

A quantum state is bi-separable if the joint probability $p(a,b,c|A,B,C)$ satisfies the condition,
\begin{equation}\label{bis}
\begin{aligned}
&p(a,b,c|A,B,C)=\sum\limits_{\lambda} p_1(\lambda) p_Q(a,b|A,B,\rho_{\lambda}^{\alpha\beta})p_Q(c|C,\rho_{\lambda}^{\gamma})\\
&\hspace{2.5cm}+\sum\limits_{\lambda} p_2(\lambda) p_Q(a|A,\rho_{\lambda}^a)p_Q(b,c|B,C,\rho_{\lambda}^{\beta\gamma})\\
&\hspace{2.5cm}+\sum\limits_{\lambda} p_3(\lambda) p(b|B,\rho_{\lambda}^b)p_Q(a,c|A,C,\rho_{\lambda}^{\alpha\gamma}),
\end{aligned}
\end{equation}
where $\sum\limits_{\lambda}p_1(\lambda)+\sum\limits_{\lambda}p_2(\lambda)
+\sum\limits_{\lambda}p_3(\lambda)=1.$
A bi-separable quantum state must not be a genuine tripartite steerable state from Alice to Bob and Charlie or from Alice and Bob to Charlie.  From (\ref{1pp}) and (\ref{GMS-AB-Cp}) a state which is not genuine tripartite steerable from Alice to Bob and Charlie is not genuine
tripartite steerable from Alice and Bob to Charlie.
As a result, given in \cite{Cavalcantin}, the noisy GHZ state demonstrates the genuine tripartite steering from Alice to Bob and Charlie in a larger region compared to that from Alice and Bob to Charlie. For general tripartite quantum states, the genuine tripartite steering from Alice and Bob to Charlie is also stronger than that from Alice to Bob and Charlie.

\

\noindent{\bf{Theorem 1}}
Let $\rho_{\mathcal{ABC}}$ be a three-qubit quantum state and
\begin{equation}\label{t1}
\begin{aligned}
\tau^1_{\mathcal{ABC}}=\mu\,\rho_{\mathcal{ABC}}
+(1-\mu)\frac{\rm{I}_2}{2}\otimes\rho_{\mathcal{BC}}
\end{aligned}
\end{equation}
with $\rho_{\mathcal{BC}}={\rm{Tr}}_{\mathcal{A}}\rho_{\mathcal{ABC}}$ and $\rm{I}_2$ the $2\times 2$ identity matrix. We have

%\begin{itemize}
(i) If $\tau^1_{\mathcal{ABC}}$ is genuine tripartite entangled, then $\rho_{\mathcal{ABC}}$ is genuine tripartite steerable from Alice to Bob and Charlie for $0\leq \mu\leq\frac{1}{\sqrt{3}}$;

(ii) If $\tau^1_{\mathcal{ABC}}$ is tripartite entangled, then $\rho_{\mathcal{ABC}}$ is tripartite steerable from Alice to Bob and Charlie for $0\leq \mu\leq\frac{1}{\sqrt{3}}.$
%\end{itemize}

The statements in Theorem 1 are equivalent to the following:

(i¡¯)  If $\rho_{\mathcal{ABC}}$ is not genuine tripartite steerable from Alice to Bob and Charlie, then $\tau^1_{\mathcal{ABC}}$ is bi-separable for $0\leq \mu\leq \frac{1}{\sqrt{3}}$;

(ii') If $\rho_{\mathcal{ABC}}$ is not tripartite steerable from Alice to Bob and Charlie, then $\tau^1_{\mathcal{ABC}}$ is fully separable for $0\leq \mu\leq\frac{1}{\sqrt{3}}$.
%\end{itemize}

\begin{proof}[Proof of Theorem 1]
We prove the theorem by proving its converse negative proposition: if $\rho_{\mathcal{ABC}}$ is not a genuine tripartite steerable state from Alice to Bob and Charlie, then $\tau^1_{\mathcal{ABC}}$ is a bi-separable state; if $\rho_{\mathcal{ABC}}$ is not a tripartite steerable state from Alice to Bob and Charlie, then
$\tau^1_{\mathcal{ABC}}$ is a fully-separable state.

Firstly we give the (unnormalized) conditional quantum state $\delta_{\mathcal{BC}}^{\beta\gamma}$ on Alice's side after Bob and Charlie perform measurements $M_B^b$ and $M_C^c$ on $\tau_{\mathcal{ABC}}^1$. Then the Bloch sphere representation of the conditional state can be expressed according to the joint
probabilities. Lastly from the condition that $\rho_{\mathcal{ABC}}$ is not genuine steering or steering from Alice to Bob and Charlie, we prove that
$\delta_{\mathcal{BC}}^{\beta\gamma}$ is the convex combination of some qubit quantum states if $\mu$ satisfies certain conditions.

{\bf{Step {1.}
}} From (\ref{t1}) we have the (unnormalized) conditional state on Alice's side when Bob and Charlie perform measurements $M_B^b$ and $M_C^c$ on $\tau_{\mathcal{ABC}}^1$,
\begin{equation}
\begin{aligned}
&\delta_{\mathcal{BC}}^{\beta\gamma}:={{\rm{Tr}}_{BC}[({\rm{I}_2}\otimes M_B^b\otimes M_C^c)}.\tau_{\mathcal{ABC}}^1]=\mu{{\rm{Tr}}_{BC}[({\rm{I}_2}\otimes M_B^b\otimes M_C^c)}.\rho_{\mathcal{ABC}}]\\ \nonumber
&\hspace{1.cm}+(1-\mu)p(b,c|B,C,\rho_{\mathcal{ABC}})\frac{{\rm{I}_2}}{2}=\frac{1}{2}(y{\rm{I}_2}+\sum\limits_{i}r_i\sigma_i),
\end{aligned}
\end{equation}
where $\sigma_i$ $(i=1,2,3)$ are Pauli matrices $\sigma_x,$ $\sigma_y$ and $\sigma_z$, respectively.

{\bf{Step 2.}} $y$ and $r_i$ $(i=1,2,3)$ are given by the joint probabilities,
\begin{equation}\nonumber
\begin{array}{rcl}
y&=&{\rm{Tr}}[\delta_{\mathcal{BC}}^{\beta\gamma}]=\mu{{\rm{Tr}}[{\rm{Tr}}_{BC}[({\rm{I}_2}\otimes M_B^b\otimes M_C^c)}.\rho_{\mathcal{ABC}}]]+(1-\mu)p(b,c|B,C,\rho_{\mathcal{ABC}})\\[2mm]
&=&p(b,c|B,C,\rho_{\mathcal{ABC}}),
\end{array}
\end{equation}
\begin{equation}\nonumber
\begin{array}{rcl}
r_i&=&{\rm{Tr}}[\delta_{\mathcal{BC}}^{\beta\gamma}.\sigma_i]={\rm{Tr}}[\delta_{\mathcal{BC}}^{\beta\gamma}.(\sigma_i^+-\sigma_i^-)]=\mu{\rm{Tr}}[{\rm{Tr}}_{\mathcal{BC}}[({\rm{I}_2}\otimes M_B^b\otimes M_C^c).\rho_{\mathcal{ABC}}].(\sigma_i^+-\sigma_i^-)]\\[2mm]
&=&\mu(p(+,b,c,|\sigma_i,B,C,\rho_{\mathcal{ABC}})-p(-,b,c,|\sigma_i,B,C,\rho_{\mathcal{ABC}})),
\end{array}
\end{equation}
with $\sigma_i^{+}$ and $\sigma_i^-$ the eigenvectors of $\sigma_i$ with respect to the eigenvalues $1$ and $-1$ of $\sigma_i$ $(i=1,2,3)$, respectively.

{\bf{Step 3.}} (I). If $\rho_{\mathcal{ABC}}$ is not a genuine tripartite steerable state from Alice to Bob and Charlie, the joint probabilities admit a hybrid LHV-LHS model as follows,
\begin{equation}
\begin{aligned}
&p(a,b,c|A,B,C)=\sum\limits_{\lambda} p_1(\lambda)p_1(a|A,\lambda)p_Q(b,c|B,C,\rho_{\lambda}^{\beta\gamma})
\\ \nonumber
&\hspace{2.5cm}+p_2(\lambda)p_Q(a,b|A,B)p_Q(c|C,\rho_{\lambda}^{\gamma})\\ \nonumber
&\hspace{2.5cm}+p_3(\lambda)p_Q(a,c|A,C)p_Q(b|B,\rho_{\lambda}^{\beta}),
\end{aligned}
\end{equation}
\begin{equation}
\begin{aligned}
&p(b,c|B,C,\rho_{\mathcal{ABC}})=\sum\limits_{\lambda}p_1(\lambda)p_{Q}(b,c|B,C,\rho_{\lambda}^{\beta\gamma})\\ \nonumber
&\hspace{2.5cm}+\sum\limits_{\lambda}p_2(\lambda)p_Q(c|C,\rho_{\lambda}^{\gamma})p_Q(b|B)\\ \nonumber
&\hspace{2.5cm}+\sum\limits_{\lambda}p_3(\lambda)p_Q(b|B,\rho_{\lambda}^{\beta})p_Q(c|C),
\end{aligned}
\end{equation}
where $p_Q(b|B)$ and $p_Q(c|C)$ are probabilities from qubit states $\rho_{\lambda}^{\beta'}={\rm{Tr}}_A[\rho_{\lambda}^{\alpha\beta}]$ and $\rho_{\lambda}^{\gamma'}={\rm{Tr}}_A[\rho_{\lambda}^{\alpha\gamma}]$ on Bob's and Charlie's sides, respectively.

\begin{equation}
\begin{aligned}
&p(\pm,b,c|\sigma_i,B,C)=\sum\limits_{\lambda} p_1(\lambda)p_1(\pm|\sigma_i,\lambda)p_Q(b,c|B,C,\rho_{\lambda}^{\beta\gamma})\\ \nonumber
&\hspace{2.5cm}+p_2(\lambda)p_Q(\pm,b|\sigma_i,B)p_Q(c|C,\rho_{\lambda}^{\gamma})\\ \nonumber
&\hspace{2.5cm}+p_3(\lambda)p_Q(\pm,c|\sigma_i,C)p_Q(b|B,\rho_{\lambda}^{\beta}).
\end{aligned}
\end{equation}

{\bf{Step 4.}} We now prove that the following conditional state $\delta_{\mathcal{BC}}^{\beta\gamma}$ is the convex combination of qubit quantum states when $\mu$ satisfies certain conditions,
\begin{equation}
\begin{aligned}
\delta_{\mathcal{BC}}^{\beta\gamma}=&\sum\limits_{\lambda}p_1(\lambda)p_Q(b,c|B,C,\rho_{\lambda}^{\beta\gamma})
\rho_{\lambda}^1+\sum\limits_{\lambda}p_2(\lambda)p_Q(c|C,\rho_{\lambda}^{\gamma})
p_Q(b|B)\rho_{\lambda}^2\\ \nonumber
&+\sum\limits_{\lambda}p_3(\lambda)p_Q(b|B,\rho_{\lambda}^{\beta})
p_Q(c|C)\rho_{\lambda}^3, \nonumber
\end{aligned}
\end{equation}
where
\begin{equation}
\begin{aligned}
&\rho_{\lambda}^1=\frac{1}{2}({\rm{I}_2}+\mu\sum\limits_i(p_1(+|\sigma_i,\lambda)
-p_1(-|\sigma_i,\lambda))\sigma_i),\\ \nonumber
&\rho_{\lambda}^2=\frac{1}{2}({\rm{I}_2}+\mu\sum\limits_i
\frac{p_Q(+,b|\sigma_i,B)-p_Q(-,b|\sigma_i,B)}{p_Q(b|B)}\sigma_i),\\ \nonumber
&\rho_{\lambda}^3=\frac{1}{2}({\rm{I}_2}+\mu\sum\limits_i
\frac{p_Q(+,c|\sigma_i,C)-p_Q(-,c|\sigma_i,C)}{p_Q(c|C)}\sigma_i). \nonumber
\end{aligned}
\end{equation}
%{\bf{Here \color{blue}{$\frac{{\rm{Tr}}[(\sigma_i\otimes M_B^b)\rho_{\lambda}^{\alpha\beta}]}{{\rm{Tr}}[({\rm{I}_2}\otimes M_B^b)\rho_{\lambda}^{\alpha\beta}]}$}is replaced by $\frac{p_Q(+,b|\sigma_i,B)-p_Q(-,b|\sigma_i,B)}{p_Q(b|B)}$ and \color{blue}{$\frac{{\rm{Tr}}[(\sigma_i\otimes M_C^c)\rho_{\lambda}^{\alpha\gamma}]}{{\rm{Tr}}[({\rm{I}_2}\otimes M_C^c)\rho_{\lambda}^{\alpha\gamma}]}$}is replaced by $\frac{p_Q(+,c|\sigma_i,C)-p_Q(-,c|\sigma_i,C)}{p_Q(c|C)}$}

Since $|p_1(+|\sigma_i,\lambda)-p_1(-|\sigma_i,\lambda)|\leq 1,$ $|\frac{p_Q(+,b|\sigma_i,B)-p_Q(-,b|\sigma_i,B)}{p_Q(b|B)}|\leq 1$ and \\
$|\frac{p_Q(+,c|\sigma_i,C)-p_Q(-,c|\sigma_i,C)}{p_Q(c|C)}|\leq 1,$ when $0\leq\mu\leq\frac{1}{\sqrt{3}}$
$\rho_{\lambda}^1,$ $\rho_{\lambda}^2$ and $\rho_{\lambda}^3$ are semi-definite positive matrices  with trace one. They are quantum states when $0\leq\mu\leq\frac{1}{\sqrt{3}}$.
Therefore,
\begin{equation}
\begin{aligned}
&p(a,b,c|A,B,C,\tau_{\mathcal{ABC}}^1)={\rm{Tr}}[M_A^a\delta_{\mathcal{BC}}^{\beta\gamma}]
=\sum\limits_{\lambda}p_1(\lambda)p_Q(b,c|B,C,\rho_{\lambda}^{\beta\gamma})p_Q(a|A,\rho_{\lambda}^1)\\ \nonumber
&\hspace{3.9cm}+\sum\limits_{\lambda}p_2(\lambda)p_Q(c|C,\rho_{\lambda}^{\gamma})p_Q(b|B)p_Q(a|A,\rho_{\lambda}^2)\\ \nonumber
&\hspace{3.9cm}+\sum\limits_{\lambda}p_3(\lambda)p_Q(b|B,\rho_{\lambda}^{\beta})p_Q(c|C)p_Q(a|A,\rho_{\lambda}^3)\\ \nonumber
&\hspace{3.5cm}=\sum\limits_{\lambda}p_1(\lambda)p_Q(b,c|B,C,\rho_{\lambda}^{\beta\gamma})p_Q(a|A,\rho_{\lambda}^1)\\ \nonumber
&\hspace{3.9cm}+\sum\limits_{\lambda}p_2(\lambda)p_Q(c|C,\rho_{\lambda}^{\gamma})p_Q(a,b|,A,B,\rho_{\lambda}^{\alpha\beta'})\\ \nonumber
&\hspace{3.9cm}+\sum\limits_{\lambda}p_3(\lambda)
p_Q(b|B,\rho_{\lambda}^{\beta})p_Q(a,c|,A,C,\rho_{\lambda}^{\alpha\gamma'}),
\end{aligned}
\end{equation}
with $p_Q(b|B)={\rm{Tr}}[M_B^b.\rho_{\lambda}^{\beta'}]$ and $p_Q(c|C)={\rm{Tr}}[M_C^c.\rho_{\lambda}^{\gamma'}],$
and
$\rho_{\lambda}^{\alpha\beta'}=\rho_{\lambda}^2\otimes\rho_{\lambda}^{\beta'}$ and
$\rho_{\lambda}^{\alpha\gamma'}=\rho_{\lambda}^3\otimes\rho_{\lambda}^{\gamma'}$.
From (\ref{bis}) $\tau_{\mathcal{ABC}}^1$ is a bi-separable state. Namely,
if $\tau^1_{\mathcal{ABC}}$ is genuine tripartite entangled, then $\rho_{\mathcal{ABC}}$ is genuine tripartite steerable from Alice to Bob and Charlie for $0\leq \mu\leq\frac{1}{\sqrt{3}}$.

{\bf{Step 3'. %MDPI: The order of step is right.
}} (II). If $\rho_{\mathcal{ABC}}$ is not tripartite steerable from Alice to Bob and Charlie, the joint probabilites admit LHV-LHS model,
\begin{equation}
\begin{aligned}
&p(a,b,c|A,B,C)
=\sum\limits_{\lambda} p(\lambda)p(a|A,\lambda)p_Q(b|B,\tau_{\lambda}^{\beta})p_Q(c|C,\tau_{\lambda}^{\gamma}),\\[1mm]\\ \nonumber
&p(b,c|B,C,\rho_{\mathcal{ABC}})=\sum\limits_{\lambda}p(\lambda)p_{Q}(b|B,\tau_{\lambda}^{b})p_Q(c|C,\tau_{\lambda}^{\gamma})
\end{aligned}
\end{equation}
and
\begin{equation}
\begin{aligned}
p(\pm,b,c|\sigma_i,B,C)=\sum\limits_{\lambda} p(\lambda)p(\pm|\sigma_i,\lambda)p_Q(b|B,\tau_{\lambda}^{\beta})p_Q(c|C,\tau_{\lambda}^{\gamma}).
\end{aligned}
\end{equation}

{\bf{Step 4'.}} Therefore, $\delta_{\mathcal{BC}}^{\beta\gamma}$ is given by the convex combination of some qubit quantum states when $\mu$ satisfies certain condition,
\begin{equation}
\begin{aligned}
&\delta_{\mathcal{BC}}^{\beta\gamma}\nonumber
=&\sum\limits_{\lambda}p(\lambda)p_Q(b|B,\tau_{\lambda}^{\beta})
p_Q(c|C,\tau_{\lambda}^{\gamma})\tau_{\lambda}^{\alpha},
\end{aligned}
\end{equation}
where
$\tau_{\lambda}^{\alpha}=\frac{1}{2}({\rm{I}_2}+\mu\sum\limits_i(p_1(+|\sigma_i,\lambda)
-p_1(-|\sigma_i,\lambda))\sigma_i).$ Since $|p_1(+|\sigma_i,\lambda)
-p_1(-|\sigma_i,\lambda|\leq 1$ for $i=1,2,3,$ when $0\leq\mu\leq\frac{1}{\sqrt{3}},$ $\tau_{\lambda}^{\alpha}$ is a semi-definite positive matrix when  $0\leq\mu\leq\frac{1}{\sqrt{3}}.$
Therefore, $\tau_{\lambda}^{\alpha}$ is a quantum state when $0\leq\mu\leq\frac{1}{\sqrt{3}}.$
Since
\begin{equation}
\begin{aligned}
p(a,b,c|A,B,C,\tau_{\mathcal{ABC}}^1)={\rm{Tr}}[M_A^a\delta_{\mathcal{BC}}^{\beta\gamma}]
=\sum\limits_{\lambda}p(\lambda)p_Q(a|A,\tau_{\lambda}^{\alpha})
p_Q(b|B,\tau_{\lambda}^{\beta})p_Q(c|C,\tau_{\lambda}^{\gamma}),\nonumber
\end{aligned}
\end{equation}
from (\ref{fulls}), $\tau_{\mathcal{ABC}}^1$ is fully separable. Hence,
if $\tau^1_{\mathcal{ABC}}$ is tripartite entangled, $\rho_{\mathcal{ABC}}$ must be tripartite steerable from Alice to Bob and Charlie for $0\leq \mu\leq\frac{1}{\sqrt{3}}$.
\end{proof}

\noindent{\bf{Theorem 2}}
Let $\rho_{\mathcal{ABC}}$ be a three-qubit state and
\begin{equation}\label{t2}
\begin{aligned}
\tau^2_{\mathcal{ABC}}=\mu\rho_{\mathcal{ABC}}+(1-\mu)\frac{\rm{I}_4}{4}\otimes\rho_{\mathcal{C}},
\end{aligned}
\end{equation}
where $\rho_{\mathcal{C}}={\rm{Tr}}_{\mathcal{AB}}\rho_{\mathcal{ABC}}$ and $\rm{I}_4$ is the $4\times 4$ identity matrix. We have\\
a) If $\tau^2_{\mathcal{ABC}}$ is genuine tripartite entangled, then $\rho_{\mathcal{ABC}}$ is genuine tripartite steerable from Alice and Bob to Charlie for $0\leq \mu\leq \frac{1}{9}$;\\
b) If $\tau^2_{\mathcal{ABC}}$ is tripartite entangled, then $\rho_{\mathcal{ABC}}$ is tripartite steerable from Alice to Bob and Charlie for $0\leq \mu\leq \frac{1}{3}$.

The proof of Theorem 2 is given in Appendix
 \ref{app1}. The statements in Theorem 2
are also equivalent to the following:

%\begin{itemize}

(a')  If $\rho_{\mathcal{ABC}}$ is not genuine tripartite steerable from Alice and Bob to Charlie, then $\tau_{\mathcal{ABC}}^2$ is bi-separable for $0\leq \mu\leq \frac{1}{9}$;

(b')  If $\rho_{\mathcal{ABC}}$ is not tripartite steerable from Alice to Bob and Charlie, then $\tau_{\mathcal{ABC}}^2$ is fully separable for $0\leq \mu\leq \frac{1}{3})$.
%\end{itemize}

%{\bf{Theorem 3}} For a quantum state $\rho_{AB}$ with dimensions $m\otimes n,$ Let
%$\tau_{AB}=\mu\rho_{\mathcal{ABC}}+(1-\mu)\frac{\rm{I}_m}{m}\otimes \rho_B,$
%if $\tau_{AB}$ is a separable state, $\rho_{AB}$ is not steerable from $A$ to $B$ for $0\leq u\leq \frac{1}{\sqrt{n^2-1}}.$

Next, we illustrate our theorems with detailed examples.\

\

\noindent{\bf{Example 1}} {Consider $\rho_{\mathcal{ABC}}=|GHZ\rangle\langle GHZ|$, where
$|GHZ\rangle=a|000\rangle+\sqrt{1-a^2}|111\rangle$. The
$\tau_{\mathcal{ABC}}^1$ defined in Theorem 1 is a $8\times 8$ matrix with entries
$\tau_{ij}$, $i,j=1,2,...,8$. The state $\tau_{\mathcal{ABC}}^1$ is genuine entangled if $|\tau_{18}|>\sqrt{\tau_{22}\tau_{77}}+\sqrt{\tau_{33}\tau_{66}}
+\sqrt{\tau_{44}\tau_{55}}$ \cite{NJP}, and
$\tau_{\mathcal{ABC}}^1$ is entangled if one of the following three inequalities is satisfied: $|\tau_{18}|>\sqrt{\tau_{22}\tau_{77}},$ $|\tau_{18}|>\sqrt{\tau_{33}\tau_{66}}$ or
$|\tau_{18}|>\sqrt{\tau_{44}\tau_{55}}$~\cite{Ting}.} Therefore,
from Theorem 1 we have that when $0<a<1$ this state $\rho_{\mathcal{ABC}}$ is tripartite steerable and also genuine tripartite steerable from Alice to Bob and Charlie.
Similarly, according to the entanglement of $\tau_{\mathcal{ABC}}^2$, from Theorem 2 we obtain that $\rho_{\mathcal{ABC}}$ is tripartitely steerable from Alice and Bob to
Charlie when $0<a<1$. While in \cite{Riccardi}, genuine tripartite steering from Alice to Bob and Charlie is detected only when  $0.5<a<0.85.$

\

%\end{Example}
\noindent{\bf{Example 2}}
Consider $\rho_{\mathcal{ABC}}=\frac{1-p}{8}\rm{I}_8+p|GHZ\rangle\langle GHZ|$ with $|GHZ\rangle=\frac{1}{\sqrt{2}}(|000\rangle+|111\rangle)$. Similar to Example 1,
by using the entanglement criteria given in \cite{NJP,Ting} and Theorem 1, we have that $\tau_{\mathcal{ABC}}^1$ is genuine tripartite entangled when $p>0.672$, and thus $\rho_{\mathcal{ABC}}$ is genuine tripartite steerable from Alice to Bob and Charlie.
When $p>0.406$ $\tau_{\mathcal{ABC}}^1$ is an entangled state, and $\rho_{\mathcal{ABC}}$ is tripartite steerable from Alice to Bob and Charlie. Furthermore, from the entanglement of $\tau_{\mathcal{ABC}}^2$ and Theorem 2, we have that $\rho_{\mathcal{ABC}}$ is tripartitely steerable from Alice and Bob to Charlie when {$p>0.6$}.
While in \cite{Riccardi} $\rho_{\mathcal{ABC}}$ is proved to be tripartite steerable {from Alice to Bob and Charlie} when $p>0.74$ and genuine tripartite steerable from Alice to Bob and Charlie when $p>0.95.$
In \cite{Jebaratnam} $\rho_{\mathcal{ABC}}$ is shown to be tripartite steerable from Alice to Bob and Charlie when $p>0.35$ and genuine steerable when $p>0.71$.
$\rho_{\mathcal{ABC}}$ is tripartite steerable from Alice and Bob to Charlie when $p>0.5$ and genuine steerable when $p>0.71$.
In \cite{Costa} $\rho_{\mathcal{ABC}}$ is shown to be tripartite steering from Alice to Bob and Charlie when $p>0.8631$ for two measurement settings, and $p>0.7642$ for three measurement settings. $\rho_{\mathcal{ABC}}$ is tripartite steering from Alice and Bob to Charlie when $p>0.6751$ for two measurement settings, and $p>0.5514$ for three measurement settings.
Hence, in the case of detecting genuine tripartite steering from Alice to Bob and Charlie, our proposed method is stronger compared with the criteria given in \cite{Jebaratnam,Riccardi,Costa},
and in the case of tripartite steering from Alice to Bob and Charlie, our proposed method is stronger with respect to the criteria in \cite{Riccardi,Costa}.
The results are listed in Table \ref{tabex2}.

%\end{Example}

\begin{table}[htbp]
\caption{Critical values to white noise p for example 2 by our theorems and the criteria in \cite{Jebaratnam,Riccardi,Costa},here S represents tripartite steering and GMS represents genuine tripartite steering}
\label{tabex2}
\begin{tabular}{|l|l|l|l|l|}
\hline
Steering  & $\mathcal{A}$ to $\mathcal{B}$ $\&$ $\mathcal{C}$(S)& $\mathcal{A}$ to $\mathcal{B}$  $\&$ $\mathcal{C}$(GMS) & $\mathcal{A}$  $\&$ $\mathcal{B}$ to $\mathcal{C}$(S) & $\mathcal{A}$  $\&$ $\mathcal{B}$ to $\mathcal{C}$(GMS)
\\ \hline
our result & 0.406  &  0.672  & 0.6  & \diagbox[width=2.8cm,height=0.32cm]{}{}\\ \hline
\cite{Jebaratnam}& 0.35  &  0.71  &  0.5& 0.71  \\ \hline
\cite{Riccardi}&  0.74  &  0.95 & \diagbox[width=2.3cm,height=0.32cm]{}{}  & \diagbox[width=2.8cm,height=0.32cm]{}{}\\ \hline
\cite{Costa}   &  0.7642  & \diagbox[width=2.8cm,height=0.32cm]{}{}  &  0.5514 & \diagbox[width=2.8cm,height=0.32cm]{}{}\\ \hline
\end{tabular}
\end{table}

Next, instead of the criteria given in \cite{NJP,Ting} we first present improved separability criteria. Consider a three-qubit state $|\psi\rangle$. Let $\sigma_{ij}$ be the entries of the matrix $\sigma=|\psi\rangle\langle \psi|$. If the state $\sigma=|\psi\rangle\langle \psi|$ is  bi-separable, we have
$|\sigma_{25}|\leq\frac{1}{2}(\sigma_{11}+\sigma_{66}),$ $|\sigma_{35}|\leq \frac{1}{2}(\sigma_{11}+\sigma_{77})$ and
$|\sigma_{23}|\leq\frac{1}{2}(\sigma_{22}+\sigma_{33})$ under the bipartition $\mathcal{A}|\mathcal{BC}$;
$|\sigma_{23}|\leq\frac{1}{2}(\sigma_{11}+\sigma_{44}),$ $|\sigma_{35}|\leq \frac{1}{2}(\sigma_{11}+\sigma_{77})$ and
$|\sigma_{25}|\leq\frac{1}{2}(\sigma_{22}+\sigma_{55})$ under the bi-partition $\mathcal{B}|\mathcal{AC}$;
$|\sigma_{23}|\leq\frac{1}{2}(\sigma_{11}+\sigma_{44}),$ $|\sigma_{25}|\leq \frac{1}{2}(\sigma_{11}+\sigma_{66})$ and
$|\sigma_{35}|\leq\frac{1}{2}(\sigma_{33}+\sigma_{55})$ under the bi-partition $\mathcal{C}|\mathcal{AB}$. Hence for any pure bi-separable quantum state $\sigma$, we have
$|\sigma_{23}|+|\sigma_{25}|+|\sigma_{35}|\leq \frac{1}{2}(2\sigma_{11}+\sigma_{44}+\sigma_{66}+\sigma_{77})
+\frac{1}{2}(\sigma_{22}+\sigma_{33}+\sigma_{55})$.
The above inequalities are also satisfied for bi-separable mixed states by the convex roof construction. Therefore, we have

{\bf{Proposition 1.}}
Let $\tau_{\mathcal{ABC}}$ be any three-qubit state and $\tau_{ij}$ the entries of the $8\times 8$ matrix $\tau_{\mathcal{ABC}}$. Then $\tau_{\mathcal{ABC}}$
is genuine tripartite entangled if
\begin{equation}\label{gmew}
\begin{aligned}
&|\tau_{23}|+|\tau_{25}|+|\tau_{35}|\\
>&\frac{1}{2}(2\tau_{11}+\tau_{44}+\tau_{66}+\tau_{77})
+\frac{1}{2}(\tau_{22}+\tau_{33}+\tau_{55}).
\end{aligned}
\end{equation}
%\end{Proposition}

\

{\noindent\bf{Example 3}}
 Let us consider now $\rho_{\mathcal{ABC}}=\frac{1-p}{8}\rm{I}_8+p|W\rangle\langle W|$ with $|W\rangle=\frac{1}{\sqrt{3}}(|001\rangle+|010\rangle+|100\rangle)$.
Using the inequality (\ref{gmew}), we have that the state $\tau_{\mathcal{ABC}}^1$ defined in Theorem 1 is genuine tripartite entangled when $p>0.816,$ whereas from the result $|\tau_{23}|+|\tau_{25}|+|\tau_{35}|>\sqrt{\tau_{11}\tau_{44}}+\sqrt{\tau_{11}
\tau_{66}}+\sqrt{\tau_{11}\tau_{77}}+\frac{1}{2}(\tau_{22}+\tau_{33}+\tau_{55})$
given in \cite{NJP, Ting}, $\tau_{\mathcal{ABC}}^1$ is
genuine tripartite entangled when $p>0.862$.
%in and $p>0.78$ by SDP in \cite{Guhne} respectively.
Hence, from Theorem 1 when $p>0.816,$ $\rho_{\mathcal{ABC}}$ is genuine tripartite steerable form Alice to Bob and Charlie. Concerning the tripartite steerability, it has been shown in \cite{Szalay} that $\tau_{\mathcal{ABC}}^1$ is tripartite entangled if $(\Gamma\otimes {\rm{I}_4})\tau_{\mathcal{ABC}}^1$ is not a positive semi-definite matrix, where $\Gamma$ is the transpose with respect to subsystems $\mathcal{A},$ $\mathcal{B}$ or $\mathcal{C}$. From this criterion we have that $\tau_{\mathcal{ABC}}^1$ is tripartite entangled when $p>0.31$, i.e., $\rho_{\mathcal{ABC}}$ is tripartite steerable form Alice to Bob and Charlie for $p>0.31$. Similarly from the $\tau_{\mathcal{ABC}}^2$ given in Theorem 2 and the criteria given %MDPI: ref42 is deleted.
 \cite{Szalay}, we have that $\rho_{\mathcal{ABC}}$ is tripartite steerable form Alice and Bob to Charlie when $p>0.621$.
While in \cite{Riccardi}, $\rho_{\mathcal{ABC}}$ is proved to be tripartitely steerable from Alice to Bob and Charlie when $p>0.85$ and no genuine tripartite steerability is detected.
In \cite{Costa}, $\rho_{\mathcal{ABC}}$ is shown to be tripartite steering from Alice to Bob and Charlie when $p>0.9814$ for two measurement settings, and $p>0.8366$ for three measurement settings. $\rho_{\mathcal{ABC}}$ is tripartite steering from Alice and Bob to Charlie when $p>0.75$ for two measurement settings, and $p>0.623$ for three measurement settings.
Hence, in the case of detecting tripartite steering and genuine tripartite steering from Alice to Bob and Charlie, our proposed method is stronger with respect to the criteria in \cite{Jebaratnam,Riccardi,Costa}.
The results are listed in Table \ref{tabex3}.

%\end{Example}
\vspace{-9pt}
\begin{table}[htbp]
\caption{Critical values to white noise p for example 3 by our theorems and the criteria in \cite{Jebaratnam,Riccardi,Costa},here S represents tripartite steering and GMS represents genuine tripartite steering}
\label{tabex3}
\begin{tabular}{|l|l|l|l|l|}
\hline
Steering  & $\mathcal{A}$ to $\mathcal{B}$ $\&$ $\mathcal{C}$(S)& $\mathcal{A}$ to $\mathcal{B}$  $\&$ $\mathcal{C}$(GMS) & $\mathcal{A}$  $\&$ $\mathcal{B}$ to $\mathcal{C}$(S) & $\mathcal{A}$  $\&$ $\mathcal{B}$ to $\mathcal{C}$(GMS)
\\ \hline
our result & 0.31  &  0.816  & 0.621  & \diagbox[width=2.8cm,height=0.32cm]{}{}\\ \hline
\cite{Jebaratnam}& \diagbox[width=2.3cm,height=0.32cm]{}{}  & \diagbox[width=2.8cm,height=0.32cm]{}{}  &  \diagbox[width=2.3cm,height=0.32cm]{}{}& \diagbox[width=2.8cm,height=0.32cm]{}{}  \\ \hline
\cite{Riccardi}&  0.85  & \diagbox[width=2.8cm,height=0.32cm]{}{}  & \diagbox[width=2.3cm,height=0.32cm]{}{}  & \diagbox[width=2.8cm,height=0.32cm]{}{}\\ \hline
\cite{Costa}   &  0.8366  &  \diagbox[width=2.8cm,height=0.32cm]{}{}  &  0.623 & \diagbox[width=2.8cm,height=0.32cm]{}{}\\ \hline
\end{tabular}
\end{table}

One point to be stressed here is that, instead of the numerical results based on a semi-definite program in \cite{ Cavalcantin}, our results are derived analytically.
For the GHZ state and W state mixed white noise, our criteria are powerful in detecting the genuine tripartite steering from Alice to Bob and Charlie. Nevertheless, the criteria can not detect any genuine tripartite steering from Alice and Bob to Charlie, which illustrates that the genuine multipartite steering from Alice and Bob to Charlie is a kind of stronger quantum correlation and some more powerful criteria are needed.

%%%%%%%%%%%%%%%%%%%%%%%%%%%%%%%%%%%%%%%%%%

%%%%%%%%%%%%%%%%%%%%%%%%%%%%%%%%%%%%%%%%%%
%\section{Discussion}

%Authors should discuss the results and how they can be interpreted from the perspective of previous studies and of the working hypotheses. The findings and their implications should be discussed in the broadest context possible. Future research directions may also be highlighted.

%%%%%%%%%%%%%%%%%%%%%%%%%%%%%%%%%%%%%%%%%%
\section{Conclusions}

The tripartite steerability and genuine tripartite steerability can be detected by
detecting the multipartite entanglement and genuine multipartite entanglement of the newly constructed state analytically. Some examples show that the criteria are powerful to detect tripartite steering from Alice to Bob and Charlie, Alice and Bob to Charlie, and genuine tripartite steering from Alice to Bob and Charlie.  Besides, we give the relationship of fully separable states, non-tripartite steerable states in a one-to-two scenario and a two-to-one scenario, bi-separable states, and non-GMS states in two scenarios. More analytical powerful criteria will be studied to detect genuine multipartite steering from Alice and Bob to Charlie in future research.

%%%%%%%%%%%%%%%%%%%%%%%%%%%%%%%%%%%%%%%%%%
%\section{Patents}

%This section is not mandatory, but may be added if there are patents resulting from the work reported in this manuscript.

%%%%%%%%%%%%%%%%%%%%%%%%%%%%%%%%%%%%%%%%%%
%\vspace{6pt}

%%%%%%%%%%%%%%%%%%%%%%%%%%%%%%%%%%%%%%%%%%
%% optional
%\supplementary{The following supporting information can be downloaded at:  \linksupplementary{s1}, Figure S1: title; Table S1: title; Video S1: title.}

\vspace{6pt}
%%%%%%%%%%%%%%%%%%%%%%%%%%%%%%%%%%%%%%%%%%
\noindent{\bf{Author Contributions:}} Writing¡ªoriginal draft, Z.C.; Writing¡ªreview and editing,
S.-M.F. All authors have read and agreed to the published version of the manuscript.

\noindent{\bf{Funding:}} This work is supported by the National Natural Science Foundation of China (NSFC) under Grants 11571313, 12071179, 12075159, and 12171044.
Beijing Natural Science Foundation (Grant No. Z190005); the Academician Innovation Platform of Hainan Province; Shenzhen Institute for Quantum Science and Engineering, Southern University of Science and Technology (No. SIQSE202001).

\noindent
{\bf{Conflicts of Interest:}} The authors declare no conflict of interest.

%%%%%%%%%%%%%%%%%%%%%%%%%%%%%%%%%%%%%%%%%%
%% Optional

%%%%%%%%%%%%%%%%%%%%%%%%%%%%%%%%%%%%%%%%%%
%% Optional
%\appendixtitles{no} % Leave argument "no" if all appendix headings stay EMPTY (then no dot is printed after "Appendix A"). If the appendix sections contain a heading then change the argument to "yes".
%\appendixstart
\appendix
\section[\appendixname~\thesection]{\label{app1}}
%\subsection[\appendixname~\thesubsection]{}
\begin{proof}[Proof of Theorem 2] The proof of Theorem 2 is similar to that of Theorem 1. We prove Theorem 2 also by proving the converse negative proposition:
if $\rho_{\mathcal{ABC}}$ is not a genuine tripartite steerable state from Alice and Bob to Charlie, then
$\tau^2_{\mathcal{ABC}}$ is a bi-separable state; if $\rho_{\mathcal{ABC}}$ is not a  tripartite steerable state from Alice and Bob to Charlie, then
$\tau^2_{\mathcal{ABC}}$ is a fully-separable state.

Firstly we give the (unnormalized) conditional quantum state $\delta_{\mathcal{C}}^{\gamma}$ on Alice's and Bob's sides after Charlie performs measurements $M_C^c$ on $\tau_{\mathcal{ABC}}^2$. Then the Bloch sphere representation of the conditional state $\delta_{\mathcal{C}}^{\gamma}$
can be expressed by the joint probabilities. Lastly from the condition that $\rho_{\mathcal{ABC}}$ is not genuine steering or steering from Alice and Bob to Charlie, the theorem is proved by proving that
$\delta_{\mathcal{C}}^{\gamma}$ is the {convex} combination of some 2-qubit quantum states when $\mu$ satisfies certain condition.

{\bf{Step {1}}}. From (\ref{t2}) we have the (unnormalized) conditional quantum state on Alice's and Bob's sides after  Charlie performs measurements $M_C^c$ on $\tau_{\mathcal{ABC}}^2$,
\begin{equation}
\begin{aligned}
\delta_{\mathcal{C}}^{\gamma}=&{\rm{Tr}_{\mathcal{C}}[(\rm{I}_4\otimes M_C^c)}\tau_{\mathcal{ABC}}^2]\\ \nonumber
=&\mu{\rm{Tr}_{\mathcal{C}}[(\rm{I}_4\otimes M_C^c)}\rho_{\mathcal{ABC}}]+(1-u)p(c|C,\rho_{\mathcal{ABC}})\frac{\rm{I}_4}{4}\\ \nonumber
=&\frac{1}{4}(x{\rm{I}_4}+\sum\limits_i a_i\sigma_i\otimes{\rm{I}_2}+{\rm{I}_2}\otimes\sum\limits_i b_i\sigma_i+\sum\limits_{ij}c_{ij}\sigma_i\otimes\sigma_j).
\end{aligned}
\end{equation}

{\bf{Step 2}}. The bloch representation are given by the joint probabilities,
$x={\rm{Tr}}[\delta_{\mathcal{C}}]=\mu{\rm{Tr}[(\rm{I}_4\otimes M_C^c)}\rho_{\mathcal{ABC}}]+(1-\mu)p(c|C,\rho_{\mathcal{ABC}})=p(c|C,\rho_{\mathcal{ABC}}),$
\begin{equation}
\begin{aligned}
a_i=&{\rm{Tr}}[((\sigma_i^+-\sigma_i^-)\otimes{\rm{I}_2}).\delta_{\mathcal{C}}^{\gamma}]\\ \nonumber
=&\mu{\rm{Tr}}[(\sigma_i^+\otimes{\rm{I}_2}\otimes M_C^c).\rho_{\mathcal{ABC}}]-\mu{\rm{Tr}}[(\sigma_i^-\otimes{\rm{I}_2}\otimes M_C^c).\rho_{\mathcal{ABC}}]\\ \nonumber
=&\mu(p(+,c|\sigma^{\mathcal{A}}_i,C,\rho_{\mathcal{ABC}})-p(-,c|\sigma_{\mathcal{A}}^i,C,\rho_{\mathcal{ABC}}))\\ \nonumber
=&\mu(2p(+,c|\sigma^{\mathcal{A}}_i,C,\rho_{\mathcal{ABC}})-p(c|C,\rho_{\mathcal{ABC}})),
\end{aligned}
\end{equation}
\begin{equation}
\begin{aligned}
b_i=&{\rm{Tr}}[({\rm{I}_2}\otimes(\sigma_i^+-\sigma_i^-)).\delta_{\mathcal{C}}^{\gamma}]\\ \nonumber
=&\mu{\rm{Tr}}[({\rm{I}_2}\otimes\sigma_i^+\otimes M_C^c).\rho_{\mathcal{ABC}}]-\mu{\rm{Tr}}[({\rm{I}_2}\otimes\sigma_i^-\otimes M_C^c).\rho_{\mathcal{ABC}}]\\ \nonumber
=&\mu(p(+,c|\sigma^{\mathcal{B}}_i,C,\rho_{\mathcal{ABC}})-p(-,c|\sigma^{\mathcal{B}}_i,C,\rho_{\mathcal{ABC}}))\\ \nonumber
=&\mu(2p(+,c|\sigma^{\mathcal{B}}_i,C,\rho_{\mathcal{ABC}})-p(c|C,\rho_{\mathcal{ABC}}))
\end{aligned}
\end{equation}
and
\begin{equation}
\begin{aligned}
c_{ij}=&{\rm{Tr}}[(\sigma_i^+-\sigma_i^-)\otimes(\sigma_j^+-\sigma_j^-)\delta_{\mathcal{C}}^{\gamma}]\\ \nonumber
=&\mu{\rm{Tr}}[((\sigma_i^+-\sigma_i^-)\otimes(\sigma_j^+-\sigma_j^-)\otimes M_C^c).\rho_{\mathcal{ABC}}]\\ \nonumber
=&\mu [p(+,+,c|\sigma^{\mathcal{A}}_i,\sigma^{\mathcal{B}}_j,C,\rho_{\mathcal{ABC}})-p(+,-,c|\sigma^{\mathcal{A}}_i,\sigma^{\mathcal{B}}_j,C,\rho_{\mathcal{ABC}})\\ \nonumber
&-p(-,+,c|\sigma^{\mathcal{A}}_i,\sigma^{\mathcal{B}}_j,C,\rho_{\mathcal{ABC}})+p(-,-,c|\sigma^{\mathcal{A}}_i,\sigma^{\mathcal{B}}_j,C,\rho_{\mathcal{ABC}})]\\ \nonumber
=&\mu(2p(+,+,c|\sigma^{\mathcal{A}}_i,\sigma^{\mathcal{B}}_j,C,\rho_{\mathcal{ABC}})+2p(-,-,c|\sigma^{\mathcal{A}}_i,\sigma^{\mathcal{B}}_j,C,\rho_{\mathcal{ABC}})-p(c|C,\rho_{\mathcal{ABC}})).
\end{aligned}
\end{equation}

Therefore,
\begin{equation}
\begin{aligned}
&\delta_{\mathcal{C}}^{\gamma}=\frac{1}{4}[p(c|C,\rho_{\mathcal{ABC}}){\rm{I}_4}+\mu\sum\limits_i
(2p(+,c|\sigma^{\mathcal{A}}_i,C,\rho_{\mathcal{ABC}})-p(c|C,\rho_{\mathcal{ABC}}))\sigma_i\otimes{\rm{I}_2}\\ \nonumber
&~~~~~~+\mu\sum\limits_i(2p(+,c|\sigma^{\mathcal{B}}_i,C,\rho_{\mathcal{ABC}})-p(c|C,\rho_{\mathcal{ABC}})){\rm{I}_2}\otimes\sigma_i\\ \nonumber
&~~~~~~+\mu\sum\limits_{ij}(2p(+,+,c|\sigma^{\mathcal{A}}_i,\sigma^{\mathcal{B}}_j,C,\rho_{\mathcal{ABC}})+2p(-,-,c|\sigma^{\mathcal{A}}_i,\sigma^{\mathcal{B}}_j,C,\rho_{\mathcal{ABC}})\\ \nonumber
&~~~~~~-p(c|C,\rho_{\mathcal{ABC}}))\sigma_i\otimes\sigma_j].
\end{aligned}
\end{equation}

{\bf{Step 3}.} {\bf{(I)}} If $\rho_{\mathcal{ABC}}$ is not genuine steerable from Alice and Bob to Charlie, we have that the joint probabilities admit a hybrid LHV-LHS model as follows,
\begin{equation}\label{p-AB-C}
\begin{aligned}
&p(a,b,c|A,B,C,\rho_{\mathcal{ABC}})=\sum\limits_{\lambda}[p_1(\lambda)p(a,b|A,B,\lambda)p_Q(c|C,\rho_{\lambda}^{\gamma})\\
&\hspace{3.5cm}+p_2(\lambda)p(a|A,\lambda)p_Q(b,c|B,C)\\
&\hspace{3.5cm}+p_3(\lambda)p(b|B,\lambda)p_Q(a,c|A,C)].
\end{aligned}
\end{equation}

Specially,
\begin{equation}\label{p-AB-C-IIC}
\begin{aligned}
&p(c|C,\rho_{\mathcal{ABC}})\\
=&\sum\limits_{\lambda}[p_1(\lambda)p_Q(c|C,\rho_{\lambda}^{\gamma})
+p_2(\lambda)P_Q(c|C,\rho_{\lambda}^{\gamma'})+p_3(\lambda)P_Q(c|C,\rho_{\lambda}^{\gamma''})]\\
\end{aligned}
\end{equation}
with $P_Q(c|C,\rho_{\lambda}^{\gamma'})$ and $P_Q(c|C,\rho_{\lambda}^{\gamma''})$ the distributions from Charlie's measurement $M_C^c$ on $\rho_{\lambda}^{\gamma'}={\rm{Tr}}_{\beta}[\rho_{\lambda}^{\beta\gamma}]$ and $\rho_{\lambda}^{\gamma''}={\rm{Tr}}_{\alpha}[\rho_{\lambda}^{\alpha\gamma}]$, respectively.

\begin{equation}\label{p-AB-C-IC}
\begin{aligned}
&p(\pm,c|\sigma_{i}^{\mathcal{A}},C,\rho_{\mathcal{ABC}})
=\sum\limits_{\lambda}[p_1(\lambda)p(\pm|\sigma_{i}^{\mathcal{A}},\lambda)
p_Q(c|C,\rho_{\lambda}^{\gamma})\\
&\hspace{3.cm}+p_2(\lambda)p(\pm|\sigma_{i}^{\mathcal{A}},\lambda)p_Q(c|C,\rho_{\lambda}^{\gamma'})\\
&\hspace{3.cm}+p_3(\lambda)p_Q(\pm,c|\sigma_i^{\mathcal{A}},C)]
\end{aligned}
\end{equation}

\begin{equation}\label{p-AB-C-IB}
\begin{aligned}
&p(\pm,c|\sigma_{i}^{\mathcal{B}},C,\rho_{\mathcal{ABC}})
=\sum\limits_{\lambda}[p_1(\lambda)p(\pm|\sigma_{i}^{\mathcal{B}},\lambda)
p_Q(c|C,\rho_{\lambda}^{\gamma})\\
&\hspace{3.cm}+p_2(\lambda)p_Q(\pm,c|\sigma_i^{\mathcal{B}},C)\\
&\hspace{3.cm}+p_3(\lambda)p(\pm|\sigma_i^{\mathcal{B}},\lambda)p_Q(c|C,,\rho_{\lambda}^{\gamma''})]
\end{aligned}
\end{equation}

and
\begin{equation}\label{p-AB-C-SSC}
\begin{aligned}
&p(\pm,\pm,c|\sigma^{\mathcal{A}}_{i},\sigma^{\mathcal{B}}_j,C)=\sum\limits_{\lambda}[p_1(\lambda)p(\pm,\pm|\sigma^{\mathcal{A}}_i,\sigma^{\mathcal{B}}_j,\lambda)
p_Q(c|C,\rho_{\lambda}^{\gamma})\\
&\hspace{3cm}+p_2(\lambda)p(\pm|\sigma^{\mathcal{A}}_i,\lambda)p_Q(\pm,c|\sigma^{\mathcal{B}}_j,C)\\
&\hspace{3cm}+p_3(\lambda)p(\pm|\sigma^{\mathcal{B}}_j,\lambda)p_Q(\pm,c|\sigma^{\mathcal{A}}_i,C)].
\end{aligned}
\end{equation}

Substituting Equations (\ref{p-AB-C-IIC})--(\ref{p-AB-C-SSC}) into the expressions of $x,$ $a_i,$ $b_i$ and $c_{ij}$ $(i,j=1,2,3)$, we have
\begin{equation}
\begin{aligned}
&x=\sum\limits_{\lambda}p_1(\lambda)p_Q(c|C,\rho_{\lambda}^{\gamma})
+p_2(\lambda)p_Q(c|C,\rho^{\gamma'}_{\lambda})+p_3(\lambda)p_Q(c|C,\rho^{\gamma''}_{\lambda}),\\ \nonumber
&a_i=\mu\sum\limits_{\lambda}[2 p_1(\lambda)p(+|\sigma^{\mathcal{A}}_i,\lambda)p_Q(c|C,\rho_{\lambda}^{\gamma})
+2p_2(\lambda)p(+|\sigma^{\mathcal{A}}_i,\lambda)p_Q(c|C,\rho_{\lambda}^{\gamma'})\\ \nonumber
&~~~~~~+2 p_3(\lambda)p_Q(+,c|\sigma^{\mathcal{A}}_i,C)\\ \nonumber
&~~~~~~-p_1(\lambda)p_Q(c|C,\rho_{\lambda}^{\gamma})
-p_2(\lambda)p_Q(c|C,\rho^{\gamma'}_{\lambda})
-p_3(\lambda)p_Q(c|C,\rho^{\gamma''}_{\lambda})],\\ \nonumber
&~~~=\mu\sum\limits_{\lambda} [p_1(\lambda)p_Q(c|C,\rho_{\lambda}^{\gamma})(2p(+|\sigma^{\mathcal{A}}_i,\lambda)-1)
+p_2(\lambda)p_Q(c|C,\rho_{\lambda}^{\gamma'})(2p(+|\sigma^{\mathcal{A}}_i)-1)
\\ \nonumber
&~~~~~~+p_3(\lambda)p_Q(c|C,\rho_{\lambda}^{\gamma''})\times(2\frac{p_Q(+,c|\sigma^{\mathcal{A}}_i,C)}
{p_Q(c|C,\rho_{\lambda}^{\gamma''})}-1)]\\
&~~~=\mu\sum\limits_{\lambda} [p_1(\lambda)p_Q(c|C,\rho_{\lambda}^{\gamma})(2p(+|\sigma^{\mathcal{A}}_i,\lambda)-1)
+p_2(\lambda)p_Q(c|C,\rho_{\lambda}^{\gamma'})(2p(+|\sigma^{\mathcal{A}}_i)-1)\\ \nonumber
&~~~~~~+p_3(\lambda)p_Q(c|C,\rho_{\lambda}^{\gamma''})\times\frac{p_Q(+,c|\sigma^{\mathcal{A}}_i,C)
-p_Q(-,c|\sigma^{\mathcal{A}}_i,C)}
{p_Q(c|C,\rho_{\lambda}^{\gamma''})}],\\
&b_i=\mu\sum\limits_{\lambda}[2p_1(\lambda)p(+|\sigma^{\mathcal{B}}_i,\lambda)
p_Q(c|C,\rho_{\lambda}^{\gamma})+2p_2(\lambda)
p_Q(+,c|\sigma^{\mathcal{B}}_i,C)\\ \nonumber
&~~~~~~+2p_3(\lambda)p(+|\sigma^{\mathcal{B}}_i,\lambda)p_Q(c|C,\rho_{\lambda}^{\gamma''})\\ \nonumber
&~~~~~~-p_1(\lambda)p_Q(c|C,\rho_{\lambda}^{\gamma})-p_2(\lambda)
p_Q(c|C,\rho^{\gamma'}_{\lambda})-p_3(\lambda)p_Q(c|C,\rho^{\gamma''}_{\lambda})]\\ \nonumber
&~~~=\mu\sum\limits_{\lambda} [p_1(\lambda)p_Q(c|C,\rho_{\lambda}^{\gamma})(2p(+|\sigma^{\mathcal{B}}_i,\lambda)-1)
+p_3(\lambda)p_Q(c|C,\rho_{\lambda}^{\gamma''})(2p(+|\sigma^{\mathcal{B}}_i,\lambda)-1)\\ \nonumber
&~~~~~~+p_2(\lambda)p_Q(c|C,\rho_{\lambda}^{\gamma'})\times\frac{p_Q(+,c|\sigma^{\mathcal{B}}_i,C)
-p_Q(-,c|\sigma^{\mathcal{B}}_i,C)}{p_Q(c|C,\rho_{\lambda}^{\gamma'})}]
\end{aligned}
\end{equation}

\begin{equation}
\begin{aligned}
&c_{ij}=\mu\sum\limits_{\lambda}\{[2 p_1(\lambda)p(+,+|\sigma^{\mathcal{A}}_i,\sigma^{\mathcal{B}}_j,\lambda)
 p_Q(c|C,\rho_{\lambda}^{\gamma})+2 p_2(\lambda)p(+|\sigma^{\mathcal{A}}_i,\lambda)\\ \nonumber
&~~~~~~\times p_Q(+,c|\sigma^{\mathcal{B}}_j,C)+2p_3(\lambda)p(+|\sigma^{\mathcal{B}}_j,\lambda)
p_Q(+,c|\sigma^{\mathcal{A}}_i,C)]\\ \nonumber
&~~~~~~+\sum\limits_{\lambda} [2p_1(\lambda)p(-,-|\sigma^{\mathcal{A}}_i,\sigma^{\mathcal{B}}_j,\lambda)
p_Q(c|C,\rho_{\lambda}^{\gamma})
+2p_2(\lambda)p(-|\sigma^{\mathcal{A}}_i,\lambda)
p_Q(-,c|\sigma^{\mathcal{B}}_j,C)\\ \nonumber
&~~~~~~+2p_3(\lambda)p(-|\sigma^{\mathcal{B}}_j,\lambda)
p_Q(-,c|\sigma^{\mathcal{A}}_i,C)]\\ \nonumber
&~~~~~~-\sum\limits_{\lambda}[p_1(\lambda)p_Q(c|C,\rho_{\lambda}^{\gamma})
+p_2(\lambda)p_Q(c|C,\rho^{\gamma'}_{\lambda})+p_3(\lambda)
p_Q(c|C,\rho^{\gamma''}_{\lambda})]\}\\
&~~~=\mu\sum\limits_{\lambda}\{ p_1(\lambda)p_Q(c|C,\rho_{\lambda}^{\gamma})
(2p(+,+|\sigma^{\mathcal{A}}_i,\sigma^{\mathcal{B}}_j,\lambda)
+2p(-,-|\sigma^{\mathcal{A}}_i,\sigma^{\mathcal{B}}_j,\lambda)-1)\\
&~~~~~~+p_2(\lambda)p_Q(c|C,\rho_{\lambda}^{\gamma'})\\ \nonumber
&~~~~~~\times [2\frac{p(+|\sigma^{\mathcal{A}}_i,\lambda)
p_Q(+,c|\sigma^{\mathcal{B}}_j,C)
+p(-|\sigma^{\mathcal{A}}_i,\lambda)
p_Q(-,c|\sigma^{\mathcal{B}}_j,C)}
{p_Q(c|C,\rho_{\lambda}^{\gamma'})}-1]\\
&~~~~~~+p_3(\lambda)p_Q(c|C,\rho_{\lambda}^{\gamma''})\\
&~~~~~~\times [2\frac{p(+|\sigma^{\mathcal{B}}_j,\lambda)
p_Q(+,c|\sigma^{\mathcal{A}}_i,C)
+p(-|\sigma^{\mathcal{B}}_j,\lambda)
p_Q(-,c|\sigma^{\mathcal{A}}_i,C)}
{p_Q(c|C,\rho_{\lambda}^{\gamma''})}-1]\}\\
&~~~=\mu\sum\limits_{\lambda}\{p_1(\lambda)p_Q(c|C,\rho_{\lambda}^{\gamma})
(2p(+,+|\sigma^{\mathcal{A}}_i,\sigma^{\mathcal{B}}_j,\lambda)
+2p(-,-|\sigma^{\mathcal{A}}_i,\sigma^{\mathcal{B}}_j,\lambda)-1)\\
&~~~~~~+p_2(\lambda)p_Q(c|C,\rho_{\lambda}^{\gamma'}) (2p(+|\sigma^{\mathcal{A}}_i)-1)
\frac{p_Q(+,c|\sigma^{\mathcal{B}}_j,C)-p_Q(+,c|\sigma^{\mathcal{B}}_j,C)}{p_Q(c|C,\rho_{\lambda}^{\gamma'})}\\
&~~~~~~+p_3(\lambda)p_Q(c|C,\rho_{\lambda}^{\gamma''})
[(2p(+|\sigma^{\mathcal{B}}_j)-1)
\frac{p_Q(+,c|\sigma^{\mathcal{A}}_i,C)
-p_Q(-,c|\sigma^{\mathcal{A}}_i,C)}
{p_Q(c|C,\rho_{\lambda}^{\gamma''})}]\}.
\end{aligned}
\end{equation}

{\bf{Step 4.}} Denoting $\Delta_1,$ $\Delta_2$ and $\Delta_3$ the terms related to $p_1(\lambda)$, $p_2(\lambda)$ and $p_3(\lambda)$ in $\delta_{\mathcal{C}}^{\gamma}$,
respectively, we have
%\begin{equation}
%\begin{aligned}
%&\delta_{\mathcal{C}}^{\gamma}=\sum\limits_{\lambda}p_1(\lambda)p_Q(c|C,\rho_{\lambda}^{\gamma})\rho_{\lambda}^{\alpha\beta}
%+\sum\limits_{\lambda}p_2(\lambda)p_Q(c|C,\rho_{\lambda}^{\gamma'})
%\rho_{\lambda}^{\alpha'}\otimes\frac{\rm{I}_2}{2}
%+\sum\limits_{\lambda}p_3(\lambda)p_Q(c|C,\rho_{\lambda}^{\gamma''})
%\rho_{\lambda}^{\alpha''}\otimes\frac{\rm{I}_2}{2}\\ \nonumber
%&~~~~~~+\frac{\mu\rm{I}_2}{4}\otimes\sum\limits_{\lambda,j}p_2(\lambda)
%(p_{Q}(+,c|\sigma_{\mathcal{B}}^j,C,\rho_{\lambda}^{\beta\gamma})
%-p_{Q}(-,c|\sigma_{\mathcal{B}}^j,C,\rho_{\lambda}^{\beta\gamma}))\sigma_j\\ \nonumber
%&~~~~~~+\frac{\mu\rm{I}_2}{4}\otimes\sum\limits_{\lambda,j}
%p_3(\lambda)(p_{Q}(c|C,\rho_{\lambda}^{\alpha\gamma})(2p(+|\sigma_{\mathcal{B}}^j)-1))\sigma_j\\ \nonumber
%&~~~~~~+\frac{\mu}{4}\sum\limits_{i,j,\lambda}p_2(\lambda)
%[(2p(+|\sigma_{\mathcal{A}}^i)-1)p_Q(+,c|\sigma_{\mathcal{B}}^j,C,\rho_{\lambda}^{\beta\gamma})
%+(2p(-|\sigma_{\mathcal{A}}^i)-1)p_Q(-,c|\sigma_{\mathcal{B}}^j,C,\rho_{\lambda}^{\beta\gamma})]\sigma_i\otimes\sigma_j\\ \nonumber
%&~~~~~~+\frac{\mu}{4}\sum\limits_{i,j,\lambda}p_3(\lambda)
%[(2p(+|\sigma_{\mathcal{B}}^j)-1)p_Q(+,c|\sigma_{\mathcal{A}}^i,C,\rho_{\lambda}^{\alpha\gamma})
%+(2p(-|\sigma_{\mathcal{B}}^j)-1)p_Q(-,c|\sigma_{\mathcal{A}}^i,C,\rho_{\lambda}^{\alpha\gamma})]\sigma_i\otimes\sigma_j\\ \nonumber
%&~~~=\Delta_1+\Delta_2+\Delta_3+\Delta_4+\Delta_5,
%\end{aligned}
%\end{equation}
\begin{equation}
\begin{aligned}
\Delta_1=&\sum\limits_{\lambda}p_1(\lambda)
p_Q(c|C,\rho_{\lambda}^{\gamma})\rho_{\lambda}^{\alpha\beta}\nonumber
\end{aligned}
\end{equation}
with
\begin{equation}
\begin{aligned}
&\rho_{\lambda}^{\alpha\beta}=\frac{1}{4}[{\rm{I}_4}
+\mu\sum\limits_i(2p(+|\sigma^{\mathcal{A}}_i,\lambda)-1)\sigma_i\otimes{\rm{I}_2}
+\mu\sum\limits_i(2p(+|\sigma^{\mathcal{B}}_i,\lambda)-1){\rm{I}_2}\otimes\sigma_i\\ \nonumber
&~~~~~~+\mu\sum\limits_{ij}(2p(+,+|\sigma^{\mathcal{A}}_i,\sigma^{\mathcal{B}}_j,\lambda)
+2p(-,-|\sigma^{\mathcal{A}}_i,\sigma^{\mathcal{B}}_j,\lambda)-1)\sigma_i\otimes\sigma_j]. \nonumber
\end{aligned}
\end{equation}
\begin{equation}
\begin{aligned}
\Delta_2=&\frac{1}{4}\sum\limits_{\lambda}p_2(\lambda)
p_Q(c|C,\rho_{\lambda}^{\gamma'})[{\rm{I}_4}+\mu\sum\limits_i(2p(+|\sigma^{\mathcal{A}}_i,\lambda)-1)
\sigma_i\otimes{\rm{I}_2}\\ \nonumber
&+\mu\sum\limits_j\frac{p_Q(+,c|\sigma_j^{\mathcal{B}},C)
-p_Q(-,c|\sigma_j^{\mathcal{B}},C)}{p_{Q}(c|C,\rho_{\lambda}^{\gamma'})}\\ \nonumber
&\times [{\rm{I}_2}+\sum\limits_i(2p(+|\sigma^{\mathcal{A}}_i)-1)\sigma_i]\otimes\sigma_j]\\ \nonumber
=&\frac{1}{4}\sum\limits_{\lambda}p_2(\lambda)p_Q(c|C,\rho_{\lambda}^{\gamma'})[{\rm{I}_4}+\mu\sum\limits_{i} (\alpha_i\sigma_i\otimes {\rm{I}_2}+\beta_i{\rm{I}_2}\otimes \sigma_i) + \mu\sum\limits_{i}\alpha_i\sigma_i\sum\limits_j \beta_j\sigma_j]\\ \nonumber
&~~~~~~\equiv \sum\limits_{\lambda} p_2(\lambda)p_Q(c|C,\rho_{\lambda}^{\gamma'})\,\Omega
\end{aligned}
\end{equation}
with $\alpha_i=2 p(+|\sigma^{\mathcal{A}}_i,\lambda)-1$ and $\beta_i=\frac{p_Q(+,c|\sigma_i^{\mathcal{B}},C)
-p_Q(-,c|\sigma_i^{\mathcal{B}},C)}{p_{Q}(c|C,\rho_{\lambda}^{\gamma'})},$ and $\Omega=\frac{1}{4}[{\rm{I}_4}+\mu\sum\limits_{i} (\alpha_i\sigma_i\otimes {\rm{I}_2}+\beta_i{\rm{I}_2}\otimes \sigma_i) + \mu\sum\limits_{i}\alpha_i\sigma_i\sum\limits_j \beta_j\sigma_j].$
\begin{equation}
\begin{aligned}
&\Delta_3=\frac{1}{4}\sum\limits_{\lambda} p_3(\lambda)p_Q(c|C,\rho_{\lambda}^{\gamma''})[{\rm{I}}_4
+\mu\sum\limits_i(\alpha_i'\sigma_i\otimes{\rm{I}_2}+{\rm{I}_2}\otimes \beta_i'\sigma_i)+\mu\sum\limits_i\alpha_i'\sigma_i\otimes\sum\limits_j\beta_j'\sigma_j]\\ \nonumber
&~~~~~~\equiv \sum\limits_{\lambda} p_3(\lambda)p_Q(c|C,\rho_{\lambda}^{\gamma''})\,\omega
\end{aligned}
\end{equation}
with $\alpha_i'=\frac{p_Q(+,c|\sigma_{\mathcal{A}}^i,C)
-p_Q(-,c|\sigma_{\mathcal{A}}^i,C)}{p_Q(c|C,\rho_{\lambda}^{\gamma''})},$ $\beta_i'=2p(+|\sigma^{\mathcal{B}}_i,\lambda)-1,$
and $\omega=\frac{1}{4}[{\rm{I}}_4
+\mu\sum\limits_i(\alpha_i'\sigma_i\otimes{\rm{I}_2}+{\rm{I}_2}\otimes \beta_i'\sigma_i)+\mu\sum\limits_i\alpha_i'\sigma_i\otimes\sum\limits_j\beta_j'\sigma_j].$
%\begin{equation}
%\begin{aligned}
%&p_Q(+,c|\sigma_i^B,C,\rho_{\lambda}^{\beta\gamma})=\rm{Tr}[(\sigma_i^{B+}\otimes M_C^c)\frac{1}{4}(\rm{I}_4+\sum\limits_k e_k\sigma_k\otimes\rm{I}_2+\rm{I}_2\otimes\sum\limits_k f_k\sigma_k+\sum\limits_{k}g_k\sigma_k\otimes\sigma_k))]\\\nonumber
%=&\frac{1}{4}(1+e_i+\sum\limits_k f_k\rm{Tr}[\sigma_kM_C^c]+g_i\rm{Tr}[M_C^c\sigma_i])
%\end{aligned}
%\end{equation}

%\begin{equation}
%\begin{aligned}
%&p_Q(-,c|\sigma_i^B,C,\rho_{\lambda}^{\beta\gamma})=\rm{Tr}[(\sigma_i^{B-}\otimes M_C^c)\frac{1}{4}(\rm{I}_4+\sum\limits_k e_k\sigma_k\otimes\rm{I}_2+\rm{I}_2\otimes\sum\limits_k %f_k\sigma_k+\sum\limits_{k}g_k\sigma_k\otimes\sigma_k))]\\\nonumber
%=&\frac{1}{4}(1-e_i+\sum\limits_k f_k\rm{Tr}[\sigma_kM_C^c]-g_i\rm{Tr}[M_C^c\sigma_i])
%\end{aligned}
%\end{equation}

{\textbf{Step 5.}} We now %MDPI: we have confirmed it is right.
 prove that $\rho_{\lambda}^{\alpha\beta}$ in $\Delta_1$, $\Omega$ in $\Delta_2$ and $\omega$ in $\Delta_3$ are quantum states when $\mu$ satisfies certain conditions.

As for $\Delta_1$, $\rho_{\lambda}^{\alpha\beta}$ can be proved to be quantum states by decomposing $1$, $2p(+|\sigma^{\mathcal{A}}_i,\lambda)-1$, $2p(+|\sigma^{\mathcal{B}}_j,\lambda)-1$ and $2p(+,+|\sigma^{\mathcal{A}}_i,\sigma^{\mathcal{B}}_j,\lambda)
+2p(-,-|\sigma^{\mathcal{A}}_i,\sigma^{\mathcal{B}}_j,\lambda)-1$ into joint probabilities that Alice and Bob perform the measurements $M_A^a$ and $M_B^b$, respectively. Noting that
\begin{equation}
\begin{aligned}
&1=\frac{1}{9}\sum\limits_{ij}
(p(+,+|\sigma^{\mathcal{A}}_i,\sigma^{\mathcal{B}}_j,\lambda)
+p(+,-|\sigma^{\mathcal{A}}_i,\sigma^{\mathcal{B}}_j,\lambda)+p(-,+|\sigma^{\mathcal{A}}_i,\sigma^{\mathcal{B}}_j,\lambda)\\ \nonumber
&
~~~~~~+p(-,-|\sigma^{\mathcal{A}}_i,\sigma^{\mathcal{B}}_j,\lambda)),\\ \nonumber
&2p(+|\sigma^{\mathcal{A}}_i,\lambda)-1\\ \nonumber
=&p(+,+|\sigma^{\mathcal{A}}_i,\sigma^{\mathcal{B}}_j,\lambda)
+p(+,-|\sigma^{\mathcal{A}}_i,\sigma^{\mathcal{B}}_j,\lambda)-p(-,+|\sigma^{\mathcal{A}}_i,\sigma^{\mathcal{B}}_j,\lambda)-p(-,-|\sigma^{\mathcal{A}}_i,\sigma^{\mathcal{B}}_j,\lambda),\\ \nonumber
&2p(+|\sigma^{\mathcal{B}}_j,\lambda)-1\\ \nonumber
=&p(+,+|\sigma^{\mathcal{A}}_i,\sigma^{\mathcal{B}}_j,\lambda)
-p(+,-|\sigma^{\mathcal{A}}_i,\sigma^{\mathcal{B}}_j,\lambda)+p(-,+|\sigma^{\mathcal{A}}_i,\sigma^{\mathcal{B}}_j,\lambda)-p(-,-|\sigma^{\mathcal{A}}_i,\sigma^{\mathcal{B}}_j,\lambda),\\ \nonumber
&2p(+,+|\sigma^{\mathcal{A}}_i,\sigma^{\mathcal{B}}_j,\lambda)
+2p(-,-|\sigma^{\mathcal{A}}_i,\sigma^{\mathcal{B}}_j,\lambda)-1\\ \nonumber
=&p(+,+|\sigma^{\mathcal{A}}_i,\sigma^{\mathcal{B}}_j,\lambda)
-p(+,-|\sigma^{\mathcal{A}}_i,\sigma^{\mathcal{B}}_j,\lambda)-p(-,+|\sigma^{\mathcal{A}}_i,\sigma^{\mathcal{B}}_j,\lambda)
+p(-,-|\sigma^{\mathcal{A}}_i,\sigma^{\mathcal{B}}_j,\lambda),
\end{aligned}
\end{equation}
we have
\begin{equation}
\begin{aligned}
\Delta_1=&\frac{1}{4}\sum\limits_{\lambda}p_1(\lambda)
p_Q(c|C,\rho_{\lambda}^{\gamma})\\ \nonumber
&\times \sum\limits_{ij}[p(+,+|\sigma^{\mathcal{A}}_i,\sigma^{\mathcal{B}}_j,\lambda)
(\frac{1}{9}{\rm{I}_4}+\mu(\frac{1}{3}\sigma_i\otimes{\rm{I}_2}
+\frac{1}{3}{\rm{I}_2}\otimes\sigma_j+\sigma_i\otimes\sigma_j))\\ \nonumber
&+ p(+,-|\sigma^{\mathcal{A}}_i,\sigma^{\mathcal{B}}_j,\lambda)(\frac{1}{9}{\rm{I}_4}
+\mu(\frac{1}{3}\sigma_i\otimes{\rm{I}_2}-\frac{1}{3}
{\rm{I}_2}\otimes\sigma_j-\sigma_i\otimes\sigma_j))\\ \nonumber
&+ p(-,+|\sigma^{\mathcal{A}}_i,\sigma^{\mathcal{B}}_j,\lambda)(\frac{1}{9}{\rm{I}_4}
+\mu(-\frac{1}{3}\sigma_i\otimes{\rm{I}_2}+\frac{1}{3}{\rm{I}_2}\otimes\sigma_j
-\sigma_i\otimes\sigma_j))\\ \nonumber
&+ p(-,-|\sigma^{\mathcal{A}}_i,\sigma^{\mathcal{B}}_j,\lambda)(\frac{1}{9}{\rm{I}_4}
+\mu(-\frac{1}{3}\sigma_i\otimes{\rm{I}_2}-\frac{1}{3}{\rm{I}_2}\otimes\sigma_j
+\sigma_i\otimes\sigma_j))].
\end{aligned}
\end{equation}

When $0\leq\mu\leq\frac{1}{9},$ the matrices $\frac{1}{9}{\rm{I}_4}+\mu(\pm\frac{1}{3}\sigma_i\otimes{\rm{I}_2}
\pm\frac{1}{3}{\rm{I}_2}\otimes\sigma_i\pm\sigma_i\otimes\sigma_i)$, $i=1,2,3$, are
semi-definite positive matrices. Hence, $\rho_{\lambda}^{\alpha\beta}$ is a quantum state shared by Alice and Bob. By direct numerical calculation $\rho_{\lambda}^{\alpha\beta}$ is a quantum state when $0\leq \mu\leq 0.23.$

$\Omega$ in $\Delta$ can be proved to be a quantum state by decomposing $\sum\limits_i \alpha_i\sigma_i,$ $\sum\limits_i \beta_i\sigma_i$ and ${\rm{I}}_4$ into the eigenvectors of  $\sum\limits_i \alpha_i\sigma_i$ and  $\sum\limits_i \beta_i\sigma_i.$
Since
\begin{equation}
\begin{aligned}
&\sum\limits_i \alpha_i\sigma_i=|\alpha_0|(|\phi\rangle_{\mathcal{A}}\langle|\phi|-|\phi^{\perp}\rangle_{\mathcal{A}}\langle|\phi^{\perp}|)\\ \nonumber
&\sum\limits_i \beta_i\sigma_i=|\beta_0|(|\psi\rangle_\mathcal{B}\langle|\psi|
-|\psi^{\perp}\rangle_{\mathcal{B}}\langle|\psi^{\perp}|)
\end{aligned}
\end{equation}
with $|\alpha_0|=\sqrt{\sum\limits_i\alpha_i^2}$,  $|\beta_0|=\sqrt{\sum\limits_i\beta_i^2}$ and ${\rm{I}_4}=(|\phi\rangle_{\mathcal{A}}\langle|\phi|
+|\phi^{\perp}\rangle_{\mathcal{A}}\langle|\phi^{\perp}|)\otimes (|\psi\rangle_{\mathcal{B}}\langle|\psi|
+|\psi^{\perp}\rangle_{\mathcal{B}}\langle|\psi^{\perp}|)$, concerning $\Delta_2$ we have
\begin{equation}
\begin{aligned}
\Delta_2=&\frac{1}{4}\sum\limits_{\lambda}p_2(\lambda)
p_Q(c|C,\rho_{\lambda}^{\gamma'})[(1+\mu(|\alpha_0\beta_0|+|\alpha_0|+|\beta_0|))
|\phi\rangle_{\mathcal{A}}\langle\phi|\otimes|\psi\rangle_{\mathcal{B}}\langle\psi|
\\ \nonumber
&+(1+\mu(-|\alpha_0\beta_0|+|\alpha_0|-|\beta_0|))|\phi\rangle_\mathcal{A}\langle\phi
|\otimes|\psi^{\perp}\rangle_B\langle\psi^{\perp}|\\ \nonumber
&+(1+\mu(-|\alpha_0\beta_0|-|\alpha_0|+|\beta_0|))
|\phi^{\perp}\rangle_{\mathcal{A}}\langle\phi^{\perp}
|\otimes|\psi\rangle_{\mathcal{B}}\langle\psi|\\ \nonumber
&+(1+\mu(|\alpha_0\beta_0|-|\alpha_0|-|\beta_0|))
|\phi^{\perp}\rangle_{\mathcal{A}}\langle\phi^{\perp}
|\otimes|\psi^{\perp}\rangle_{\mathcal{B}}\langle\psi^{\perp}|]\\ \nonumber
\equiv & \sum\limits_{\lambda}p_2(\lambda) p_Q(c|C,\rho_{\lambda}^{\gamma'})\, \Omega.
\end{aligned}
\end{equation}

Let $q_1(\lambda)=\frac{1}{4}(1+\mu(|\alpha_0\beta_0|+|\alpha_0|+|\beta_0|)),$ $q_2(\lambda)=\frac{1}{4}(1+\mu(-|\alpha_0\beta_0|+|\alpha_0|-|\beta_0|)),$ $q_3(\lambda)=\frac{1}{4}(1+\mu(-|\alpha_0\beta_0|-|\alpha_0|+|\beta_0|))$ and $q_4(\lambda)=\frac{1}{4}(1+\mu(|\alpha_0\beta_0|-|\alpha_0|-|\beta_0|)), \rho_{\lambda}^{\alpha,i}=|\phi\rangle_\mathcal{A}\langle\phi
|$ or $|\phi^{\perp}\rangle_{\mathcal{A}}\langle\phi^{\perp}
|$ and $\rho_{\lambda}^{\beta',i}=|\psi\rangle_{\mathcal{B}}\langle\psi|$ or $|\psi^{\perp}\rangle_B\langle\psi^{\perp}|.$
Since $|\alpha_i|\leq 1$ and $|\beta_i|\leq 1$ $(i=1,2,3),$ $|\alpha_0|\leq \sqrt{3}$ and $|\beta_0|\leq \sqrt{3}$, the maximum value of $q_1(\lambda)$ is $\frac{1}{4}(1+\mu(3+2\sqrt{3})),$ the minimum values of $q_2(\lambda)$ and $q_3(\lambda)$ are $\frac{1}{4}(1-3\mu)$ and the minimum value of  $q_4(\lambda)$ is $\frac{1}{4}(1-\sqrt{3}\mu).$
One verifies that when $0\leq \mu\leq \frac{1}{3}$, the coefficients $q_i(\lambda)$ $(i=1,\cdots,4)$ are all positive, and their summation is one. The matrix $\Omega$
is a quantum state with the first subsystem determined by party $\mathcal{A}$ and
the second subsystem determined by parties $\mathcal{B}$ and $\mathcal{C}$. Denote $\Omega=\sum\limits_{i}q_i(\lambda)\rho_{\lambda}^{\alpha,i}\otimes
\rho_{\lambda}^{\beta',i}$ with $\sum\limits_{i}q_i(\lambda)=1$.

Similarly, we can prove that
\begin{equation}
\begin{aligned}
&\Delta_3=\frac{1}{4}\sum\limits_{\lambda} p_3(\lambda)p_Q(c|C,\rho_{\lambda}^{\gamma''})[{\rm{I}_4}
+\mu\sum\limits_i(\alpha_i'\sigma_i\otimes{\rm{I}_2}+{\rm{I}_2}\otimes \beta_i'\sigma_i)+\mu\sum\limits_i\alpha_i'\sigma_i\otimes\sum\limits_j\beta_j'\sigma_j]\\ \nonumber
&~~~~~~\equiv \sum\limits_{\lambda} p_3(\lambda)p_Q(c|C,\rho_{\lambda}^{\gamma''})\,\omega
\end{aligned}
\end{equation}
with $\alpha_i'=\frac{p_Q(+,c|\sigma^{\mathcal{A}}_i,C)
-p_Q(-,c|\sigma^{\mathcal{A}}_i,C)}
{p_Q(c|C,\rho_{\lambda}^{\gamma''})}$
and $\beta_i'=2p(+|\sigma^{\mathcal{B}}_i,\lambda)-1.$ When $0\leq \mu\leq \frac{1}{3}$ we verify that $\omega$ is a quantum state determined by parties $\mathcal{A}$, $\mathcal{C}$ and $\mathcal{B}.$
We denote $\omega=\sum\limits_{i}q'_i(\lambda)\rho_{\lambda}^{\alpha',i}\otimes\rho_{\lambda}^{\beta,i}$ satisfying $\sum\limits_{i}q'_i(\lambda)=1$.

Therefore,
\begin{equation}
\begin{aligned}
p(a,b,c|M_A,M_B,M_C)=&{\rm{Tr}}[M_A^a\otimes M_B^b\otimes M_C^c.\tau_{\mathcal{ABC}}^2]={\rm{Tr}}[M_A^a\otimes M_B^b.\delta_C^c]\\ \nonumber
=&\sum\limits_{\lambda}p_1(\lambda)p_Q(a,b|A,B,\rho_{\lambda}^{\alpha\beta})p_Q(c|C,\rho_{\lambda}^{\gamma})\\ \nonumber
&+\sum\limits_{\lambda'}p'_2(\lambda')p_Q(a|A,\rho_{\lambda'}^{\alpha})p_Q(b,c|B,C,\rho_{\lambda'}^{\beta\gamma'})\\ \nonumber
&+\sum\limits_{\lambda''}p'_3(\lambda'')p_Q(a,c|A,C,\rho_{\lambda''}^{\alpha\gamma'})
p_Q(b|B,\rho_{\lambda''}^{\beta})
\end{aligned}
\end{equation}
{with $p'_2(\lambda')=p_2(\lambda)q_i(\lambda),$ $p_Q(b,c|B,C,\rho_{\lambda'}^{\beta\gamma'})=p_Q(c|C,\lambda,\rho_{\lambda}^{\gamma'})
p_Q(b|B,\rho_{\lambda}^{\beta',i}),$ $\rho_{\lambda'}^{\alpha}=\rho_{\lambda}^{\alpha,i}$
and  $p'_3(\lambda'')=p_3(\lambda)q_i'(\lambda),$ $p_Q(a,c|A,C,\rho_{\lambda''}^{\alpha\gamma'})=p_Q(c|C,\lambda,\rho_{\lambda}^{\gamma''})
p_Q(a|A,\rho_{\lambda}^{\alpha',i}),$ $\rho_{\lambda''}^{\beta}=\rho_{\lambda}^{\beta,i}.$
Since $\sum\limits_{\lambda}q_i(\lambda)=1$ and $\sum\limits_{\lambda}q_i'(\lambda)=1,$  $\sum\limits_{\lambda} p_1(\lambda)+\sum\limits_{\lambda'}p_2'(\lambda')+\sum\limits_{\lambda''}p_3'(\lambda'')=1.$}
Namely, if $\rho_{\mathcal{ABC}}$ is not genuine tripartite steerable from Alice and Bob to Charlie, then $\tau_{\mathcal{ABC}}^2$ is bi-separable for $0\leq \mu\leq \frac{1}{9}$ analytically.

{\bf{Step{ 3'.} %MDPI: we have confirmed that the order of steps are right.
}} {\bf{(II)}} We next prove that $\delta_{\mathcal{C}}^{\gamma}$ is the convex combination of some quantum states when $\mu$ satisfies certain conditions for tripartite steering from Alice and Bob to Charlie.
If $\rho_{\mathcal{ABC}}$ is not tripartite steerable from Alice and Bob to Charlie, we have the joint probabilities admiting LHV-LHS models,
\begin{equation}\label{p-AB-C-p}
\begin{aligned}
&p(a,b,c|A,B,C)=\sum\limits_{\lambda}p(\lambda)p(a|A,\lambda)p(b|B,\lambda)
p_Q(c|C,\tau_{\lambda}^{\gamma}).
\end{aligned}
\end{equation}

Specially,
\begin{equation}\label{p-AB-C-IICp}
\begin{aligned}
&p(c|C)=\sum\limits_{\lambda}p(\lambda)p_Q(c|C,\tau_{\lambda}^{\gamma}),
\end{aligned}
\end{equation}

\begin{equation}\label{p-AB-C-ICp}
\begin{aligned}
&p(\pm,c|\sigma^{\mathcal{A}}_i,C)=\sum\limits_{\lambda}p(\lambda)p(\pm|\sigma^{\mathcal{A}}_i,\lambda)p_Q(c|C,\tau_{\lambda}^{\gamma}),
\end{aligned}
\end{equation}
\begin{equation}\label{p-AB-C-ICpp}
\begin{aligned}
&p(\pm,c|\sigma^{\mathcal{B}}_j,C)=\sum\limits_{\lambda}p(\lambda)p(\pm|\sigma^{\mathcal{B}}_j,\lambda)p_Q(c|C,\tau_{\lambda}^{\gamma}),
\end{aligned}
\end{equation}
and
\begin{equation}\label{p-AB-C-SSCp}
\begin{aligned}
&p(\pm,\pm,c|\sigma^{\mathcal{A}}_i,\sigma^{\mathcal{B}}_j,C)=\sum\limits_{\lambda}p(\lambda)p(\pm|\sigma^{\mathcal{A}}_i,\lambda)p(\pm|\sigma^{\mathcal{B}}_j,\lambda)
p_Q(c|C,\tau_{\lambda}^{\gamma})
\end{aligned}
\end{equation}

Substituting Equations (\ref{p-AB-C-IICp})--(\ref{p-AB-C-SSCp}) into the expressions of $x$, $a_i,$ $b_i$ and $c_{ij}(i=1,2,3$, \mbox{$j=1,2,3),$} we have
\begin{equation}
\begin{aligned}
&x=\sum\limits_{\lambda}p(\lambda)p_Q(c|C,\tau_{\lambda}^{\gamma}),\\ \nonumber
&a_i=\mu\sum\limits_{\lambda}[2 p(\lambda)p(+|\sigma^{\mathcal{A}}_i)p_Q(c|C,\tau_{\lambda}^c)-p(\lambda)p_Q(c|C,\tau_{\lambda}^{\gamma})],\\ \nonumber
&b_i=\mu\sum\limits_{\lambda}[2 p(\lambda)p(+|\sigma^{\mathcal{B}}_i)p_Q(c|C,\tau_{\lambda}^c)-p(\lambda)p_Q(c|C,\tau_{\lambda}^{\gamma})],\\ \nonumber
&c_{ij}=\mu\sum\limits_{\lambda}[2 p(\lambda)p(+|\sigma^{\mathcal{A}}_i,\lambda)p(+|\sigma_{\mathcal{B}}^j,\lambda)p_Q(c|C,\tau_{\lambda}^{\gamma})\\ \nonumber
&~~~~~~+2p(\lambda)p(-|\sigma^{\mathcal{A}}_i,\lambda)p(-|\sigma^{\mathcal{B}}_j,\lambda)p_Q(c|C,\tau_{\lambda}^{\gamma})
-p(\lambda)p_Q(c|C,\rho_{\lambda}^{\gamma})].
\end{aligned}
\end{equation}

Therefore,
\begin{equation}
\begin{aligned}
&\delta_{\mathcal{C}}^{\gamma}=\frac{1}{4}\sum\limits_{\lambda}p(\lambda)
p_Q(c|C,\tau_{\lambda}^c)({\rm{I}_4}+\mu\sum\limits_i(2p(+|\sigma^{\mathcal{A}}_i)-1)
\sigma_i\otimes{\rm{I}_2}\\ \nonumber
&~~~~~~+\mu\sum\limits_i(2p(+|\sigma^{\mathcal{B}}_i)-1){\rm{I}_2}\otimes\sigma_i\\ \nonumber
&~~~~~~+\mu\sum\limits_{ij}(2p(+|\sigma^{\mathcal{A}}_i)p(+|\sigma^{\mathcal{B}}_j)+2p(-|\sigma^{\mathcal{A}}_i)
p(-|\sigma^{\mathcal{B}}_j)-1)\sigma_i\otimes\sigma_j).
\end{aligned}
\end{equation}

Since
\begin{equation}
\begin{aligned}
&\sum\limits_{ij}2p(+|\sigma^{\mathcal{A}}_i,\lambda)p(+|\sigma^{\mathcal{B}}_j,\lambda)+2p(-|\sigma^{\mathcal{A}}_i,\lambda)
p(-|\sigma^{\mathcal{B}}_j,\lambda)-1\\ \nonumber
=&\sum\limits_i(2p(+|\sigma^{\mathcal{A}}_i,\lambda)-1)
\sum\limits_j(2p(+|\sigma^{\mathcal{B}}_j,\lambda)-1),    \nonumber
\end{aligned}
\end{equation}
$\delta_{\mathcal{C}}^{\gamma}$ can be written as
\begin{equation}
\begin{aligned}
&\delta_{\mathcal{C}}^{\gamma}=p(\lambda)p_Q(c|C,\tau_{\lambda}^c)\chi,
\end{aligned}
\end{equation}
where $\chi=\frac{1}{4}[{\rm{I}}_4+\sum\limits_{i}a''_i\sigma_i\otimes {\rm{I}_2}+\sum\limits_{i}b''_i\sigma_i {\rm{I}_2}\otimes \sigma_i+\sum\limits_{i}a''_i\sigma_i\otimes {\rm{I}_2}\sum\limits_{j}b^{''}_j\sigma_j {\rm{I}_2}\otimes \sigma_i]$ with \linebreak  $a_i^{''}=2p(+|\sigma^{\mathcal{A}}_i)-1$ and $b_i^{''}=2p(+|\sigma^{\mathcal{B}}_i)-1 (i=1,2,3).$
With {$\Delta_2$ and $\Delta_3$ in {Step {5} %MDPI: we have confirmed
}}, we can prove that $\delta_C^c$ is a quantum state when $0\leq \mu\leq \frac{1}{3}$, which implies that $\tau_{\mathcal{ABC}}^2$ is fully separable.
Namely, if $\rho_{\mathcal{ABC}}$ is not tripartite steerable from Alice to Bob and Charlie, then $\tau_{\mathcal{ABC}}^2$ is fully separable for $0\leq \mu\leq \frac{1}{3})$.
\end{proof}

%\section[\appendixname~\thesection]{}
%All appendix sections must be cited in the main text. In the appendices, Figures, Tables, etc. should be labeled, starting with ``A''---e.g., Figure A1, Figure A2, etc.

%%%%%%%%%%%%%%%%%%%%%%%%%%%%%%%%%%%%%%%%%%


%\reftitle{References}

% Please provide either the correct journal abbreviation (e.g. according to the ¡°List of Title Word Abbreviations¡± http://www.issn.org/services/online-services/access-to-the-ltwa/) or the full name of the journal.

%\bibliography{your_external_BibTeX_file}

%=====================================
% References, variant B: internal bibliography
%=====================================

\end{document}